\documentclass[aps,
showpacs,preprintnumbers,amsmath,amssymb,nofootinbib,longbibliography]{revtex4-2}
\usepackage[colorlinks=true, pdfstartview=FitV, linkcolor=red, citecolor=blue, urlcolor=black, pdftitle={},pdfauthor={},pdfsubject={}, pdfkeywords={}]{hyperref}
\usepackage[dvipdfmx]{graphicx} 
\usepackage{amsmath,amssymb,bm}
\usepackage{mathtools}
\usepackage{color}
\usepackage{physics}
\usepackage{comment}

\usepackage{colortbl}

\newcommand{\vp}{\varphi}

\newcommand{\vs}{\varsigma}

\newcommand{\vrho}{\varrho}

\newcommand{\cd}{\! \cdot \!}

\newcommand{\Ra}{\rangle}
\newcommand{\La}{\langle}
\newcommand{\p}{\partial}

\newcommand{\mf}[1]{\mathfrak{ #1}}
\newcommand{\cl}[1]{\mathcal{ #1}}
\newcommand{\tx}[1]{\mathrm{ #1}}

\newcommand{\eqn}[1]{\begin{equation}\begin{split}#1\end{split}\end{equation}}

\begin{document}

\preprint{YITP-23-95}
\author{Sinya Aoki}
\email{saoki@yukawa.kyoto-u.ac.jp}
\affiliation{
Center for Gravitational Physics and Quantum Information, Yukawa Institute for Theoretical Physics, Kyoto University, Kitashirakawa Oiwakecho, Sakyo-Ku, Kyoto 606-8502, Japan
}
\author{J\'{a}nos Balog}
\email{balog.janos@wigner.hu}
\affiliation{
Holographic QFT Group, Institute for Particle and Nuclear Physics, Wigner Research Centre for Physics, H-1525 Budapest 114, P.O.B. 49, Hungary
}
\author{Kiyoharu Kawana}
\email{kkiyoharu@kias.re.kr}
\affiliation{
School of Physics, Korea Institute for Advanced Study, Seoul 02455, Korea
}
\author{Kengo Shimada}
\email{kengo.shimada@yukawa.kyoto-u.ac.jp}
\affiliation{
Center for Gravitational Physics and Quantum Information, Yukawa Institute for Theoretical Physics, Kyoto University, Kitashirakawa Oiwakecho, Sakyo-Ku, Kyoto 606-8502, Japan
}


\title{Bulk modified gravity from a thermal CFT\\ by the conformal flow}
\begin{abstract}
We construct a bulk spacetime from a boundary CFT, $O(N)$ free scalar model, at finite temperature using a smearing technique, called  a conformal flow.
The bulk metric is constructed as an information metric associated with the boundary thermal state.
Near the boundary (UV region), an asymptotically AdS spacetime is obtained with a leading order perturbation of scalar mode.
Based on the falloff behavior of the perturbations and the $O(N)$ symmetry in the CFT,
we argue that the corresponding bulk theory is a modified gravity with scalar mode such as $f(R)$ gravity rather than Einstein's general relativity coupled minimally to matter fields.
Moving to Einstein frame, we show that the metric is asymptotically the same as the AdS black brane solution.
On the other hand, deep in the bulk (IR region), the spacetime turns out to be conformally equivalent to the near horizon limit of AdS extremal black brane, though it is no longer a solution of $f(R)$ gravity, and hence more general classes of modified gravity need to be considered.

\end{abstract}
\maketitle


\section{Introduction}
The AdS/CFT correspondence \cite{Maldacena:1997re,Gubser:1998bc,Witten:1998qj} tells us that any excitation on the ($d+1$)-dimensional anti-de Sitter (AdS) spacetime has a dual description with a operator on the $d$-dimensional conformal field theory (CFT).
Although it is conjectured based on a specific duality associated with D-branes in superstring theory,
a generic CFT is expected to have a dual theory on the AdS because of the isomorphism between the symmetry groups of the two theories.

This isomorphism   motivates ones to explore the possibility of constructing local bulk field $\hat{\Phi}$ by smearing of a boundary single trace operator $\hat{S}$ \cite{Banks:1998dd} where
\eqn{
\lim_{z\to 0} \hat{\Phi} =  z^{\Delta_S}  \times  \hat{S} \label{BDHM}
}
holds in the AdS boundary limit characterized by $z\to 0$ with $\Delta_S$ being conformal dimension of $\hat{S}$.
An explicit construction is given by Hamilton, Kabat, Lifschytz, and Lowe (HKLL) for the large $N$ limit where the bulk fields become free \cite{Hamilton:2005ju,Hamilton:2006az
}; interacting cases are also studied in \cite{Kabat:2011rz
}.
In this approach, since the operators to be smeared are singlet, their conformal dimension is implicitly assumed to be greater than $d-1$ in the original papers; and later in \cite{DelGrosso:2019gow}, the range of allowed conformal dimension was extended down to $d/2$ by analytic continuation, see also \cite{Aoki:2021ekk,Aoki:2023lgr}.

\ 

There exists a different approach to the bulk reconstruction using a smearing method with flow equation \cite{Aoki:2015dla,Aoki:2016ohw,Aoki:2017bru,Aoki:2017uce,Aoki:2018dmc,Aoki:2022lye}, where the elementary field rather than the singlet operators in the boundary theory is smeared.
The bulk spacetime is constructed from the CFT in such a way that the metric can be interpreted as an information metric associated with a state of the boundary \cite{Aoki:2018dmc}.
This approach manifests the emergence of the bulk spacetime itself, including the AdS background.
In \cite{Aoki:2022lye}, a special form of the flow equation is proposed to reproduce the relation that summarizes the AdS/CFT correspondence, conjectured by Gubser, Klebanov, Polyakov \cite{Gubser:1998bc} and Witten \cite{Witten:1998qj} (GKP-Witten).
The use of this special flow equation may make it possible to discuss what is essential to the correspondence from  a different perspective in the flow equation approach; the bulk reconstruction is based on the smeared elementary field whose conformal dimension is less than $d/2$, and thus, the flow approach is complementary to the HKLL one.

The flow method has been applied to a case with a thermal state of the boundary CFT in \cite{Aoki:2020ztd} and the resultant bulk spacetime was compared with the AdS black brane, which is the expected dual to the thermalized CFT in the standard AdS/CFT scenario.
However, in that work, a conventional (the simplest) flow equation was employed and in addition,  the definition of the bulk metric lacks the information-theoretical interpretation.
Besides, the comparison with the black brane is made in a rather qualitative way.
In the present paper, we therefore revisit the same setup but with the special flow equation proposed in \cite{Aoki:2022lye} and define the bulk geometry by the information metric to compare its asymptotic behavior with the AdS black brane.

A short summary of results in this paper is as follows.
(1) The bulk space constructed to the thermal state becomes the asymptotic AdS in $d+1$ dimensions near boundary (UV), while it approaches to the AdS space in $d$ dimensions deep in the bulk (IR)
in accordance with the dimensional reduction of boundary CFT at high temperature.
(2) More precisely, the metric in the IR region describes AdS$_2 \times$ R$_{d-1}$.
(3) The leading deviations of the metric from the AdS in the UV region contain both scalar and tensor modes, which satisfies the equations of motion for the $f(R)$ gravity.
(4) The metric deviations in the UV region describes the black brane solution in the Einstein frame.
(5) Relations between bulk modes and  boundary operator similar to eq.~(\ref{BDHM}) are derived for scalar and tensor modes.

\ 

This paper is organized as follows.
In Sec.\ref{Sec-Bulk reconstruction by flow equation},
we briefly review the flow method for the boundary CFT in the vacuum state. 
We show how the bulk metric is obtained from the flowed elementary field and how it is interpreted as a quantum information metric. 
It is also discussed that the GKP-Witten relation is constructed  for a special choice of the flow equation, which convert the boundary conformal symmetry to the bulk AdS isometry.
In Sec.\ref{Sec-Bulk geometry from thermal boundary state},
we generalize the formulation to the case with the thermal CFT.
We normalized the flowed field in such a way that the bulk metric can be interpreted as a quantum information metric even for the thermal state. We then explicitly compute the bulk metric
and discuss its asymptotic behaviors both in UV and IR.
In Sec.\ref{Sec-Correspondence between boundary and bulk},  we focus on the boundary-bulk correspondence in the UV region.
We show that the bulk theory is consistent with the $f(R)$ gravity, and
the metric perturbations in the Einstein frame reproduce those of the AdS black brane.
We also extract explicit relations between propagating modes in the bulk and operators in the boundary CFT.
We conclude our paper in Sec.\ref{Sec-Summary}.
Some technical details are presented in Appendices.

\section{Bulk reconstruction by flow equations and the conformal flow\label{Sec-Bulk reconstruction by flow equation}}
In this section, we review the bulk reconstruction from a boundary CFT with a smearing technique based on a flow equation.
This method does not assume an existence of the AdS spacetime a priori.
The bulk metric is constructed from the smeared CFT fields, and can be interpreted as a quantum information metric for a mixed state.
As discussed below, with a certain choice of the flow equation, the GKP-Witten relation is reproduced.

As a boundary CFT, we consider a $O(N)$ scalar field $\hat{\vp}^a (x)$ with $a= 1,2,\cdots,N$ in a $d$-dimensional Euclidean spacetime. In this review part, we do not specify the corresponding action,  but only assume that a 2-pt function behaves as
\eqn{
\La \hat{\vp}^a (x)\hat{\vp}^b (x') \Ra_0 = \delta^{ab} \frac{C_0}{\abs{x-x'}^{2 \Delta}} =\delta^{ab}\omega \int {\dd^d p \over (2\pi)^d} {e^{i p(x-x')} \over (p^2)^{\bar\Delta} }~,
\label{vacuum-two-point-func.}
}
where $C_0$ being a constant, $\Delta$ is a conformal dimension of $\hat\vp$, and 
\eqn{\omega:= C_0 {\Gamma(\bar\Delta) 2^{2\bar\Delta}\pi^{d\over 2}\over \Gamma(\Delta)}, \quad
\bar\Delta:= {d\over 2}-\Delta.
}
The bracket with subscript $0$ denotes an expectation value in the vacuum state:
$\La \hat{\cl{O}} \Ra_0 := \bra{0} \hat{\cl{O}} \ket{0}$.
The integer $N$ should be large to justify a classical picture of a bulk curved spacetime.

\subsection{Emergent dimension with smearing \label{Sec-Smearing}}
A flowed field operator $\hat{\phi}^a (x;\eta)$ is introduced as an extension of 
the boundary operator $\hat\vp^a$ with respect to a ``flow time'' $\eta \ge  0$ to satisfy a flow equation as
\eqn{
(-\alpha \eta \p_\eta^2 +\beta \p_\eta) \hat{\phi}^a(x;\eta) = \p^2 \hat{\phi}^a(x;\eta), \qquad
\hat{\phi}^a(x;0) = \hat{\vp}^a(x), 
\label{flow_general}
}
where $\alpha \ge 0$ and $\beta$ are real parameters.
A solution to the above flow equation is written in an integral form as
\eqn{
\hat{\phi}^a(x;\eta) = \int \dd^dy\, S(x-y;\eta) \hat\vp^a(y)=
\int {{\dd^dp}\over (2\pi)^d} e^{i px}\tilde S(p;\eta) \hat{\tilde{\vp}}^a(p),
}
which may be regarded as a smearing of $\vp^a$ with a kernel $S$.
Indeed, for a simplest choice that $\alpha=0$ and $\beta=1$, the kernel represents a Gaussian smearing as
\eqn{
S(x;\eta) ={1\over (4\pi\eta)^{d\over 2} } \exp{-\abs{x}^2/4\eta} ~.
\label{simplest choice}
}
A kernel for a solution to the flow equation \eqref{flow_general}  can be obtained 
in the Fourier space as
\eqn{
\tilde S(p;\eta)= {2\over \Gamma(\nu)}\left({\sqrt{\eta\over \alpha}} p\right)^\nu K_\nu\left( 2{\sqrt{\eta\over \alpha}} p \right) ~,
}
where $\nu:= 1+\beta /\alpha$, and $K_\nu$ is the modified Bessel function of the second kind, 
which leads to
\eqn{
S(x;\eta)=
{\Gamma(\nu+{d\over 2})\over \Gamma(\nu)}\left({\alpha\over 4\pi\eta}\right)^{d\over 2}\left( 1+{\alpha x^2\over 4\eta}\right)^{-\nu-{d\over2}}~.
\label{general choice}
}
Thanks to the smearing by the flow equation, operators with $\eta >0$ do not have any contact singularities.
Note that $\eta$'s mass dimension is $-2$ and it turns out that $1/\sqrt{\eta}$ plays the role of UV cutoff. 
In this sense,  we refer to $\eta \to 0$ as the UV limit and $\eta \to \infty$ as the IR limit.

Let us introduce a normalized field operator
\eqn{
\hat{\sigma}_0^a(X) := \frac{\hat{\phi}^a(x;\eta)}{\sqrt{ \La\hat{\phi}^2(x;\eta)\Ra_0}} ~, \quad
\hat{\phi}^2(x;\eta):= \sum_{a=1}^N\hat{\phi}^a(x;\eta)\hat{\phi}^a(x;\eta)
\label{def-sigma_0}
}
where $X^M = (x^\mu , z)$ with $z\propto \sqrt{\eta}$.\footnote{The Greek letters like $\mu, \nu, \sigma$ run from $0$ to $d-1$ and the lowercase Latin letters like $i, j$ run from $1$ to $d-1$. The Euclidean time coordinate is denoted by $\tau := x^0$. The uppercase Latin letters like $M, N$ running from $0$ to $d$ are used for $d+1$-dimensional Euclidean spacetime including the emergent direction.}
By definition, $\La \sigma^a(X) \sigma^b(X)\Ra_0 =\delta^{ab}/N$.
We then define a bulk metric operator
\eqn{
\hat{g}_{MN} (X) := \ell^2 \times \sum_{a=1}^N \frac{\p \hat{\sigma}_0^a(X)}{\p X^M} \frac{\p \hat{\sigma}_0^a(X)}{\p X^N} ~, \label{ghat^vac}
}
 where  $\ell$ is an arbitrary length scale.
A vacuum expectation value (VEV) of the metric operator, 
\eqn{
g^\tx{vac}_{MN}(X) := \La \hat{g}_{MN} (X) \Ra_0  = \ell^2 \times \eval{\frac{\p}{\p X^M} \frac{\p}{\p X'^N} G_0(X,X') }_{X=X'}
\label{g^vac}
}
with the two-point function
\eqn{
 G_0(X,X') := \sum_{a=1}^N \La \hat{\sigma}_0^a(x;\eta) \hat{\sigma}_0^a(x';\eta') \Ra_0  ~,
 \label{def-G_0}
}
can be interpreted as the Bures information metric \cite{Uhlmann1992} in the following way. 
In general, an infinitesimal distance between mixed density operators is defined by
\eqn{
\dd^2(\rho_0,\rho_0+\dd \rho_0):={1\over 2}\Tr{d\rho_0 \hat G}~,
}
where an operator $\hat G$ satisfies $ \rho_0 \hat G +\hat G\rho_0 =\dd \rho_0$. In the present case, 
the mixed density operator is given by\cite{Aoki:2017bru}
\eqn{
\rho_0(X):=& \sum_{a=1}^N \hat{\sigma}^a(x;\eta)\ket{0}\bra{0} \hat{\sigma}^a(x;\eta) ~, 
\quad
\rho_0(X) \rho_0(X) = \rho_0(X) /N~, 
\label{vacuum}
}
from which we read off $\hat{G}= N \dd \rho_0(X) = N \dd X^M \p_M \rho_0(X) $. Therefore,
the distance
\eqn{
\frac{1}{2} \tx{Tr} \qty{\dd \rho_0(X) \hat{G}} = \frac{N}{2} \tx{Tr} \qty{\p_M \rho_0(X) \p_N \rho_0(X) } \dd X^M \dd X^N \label{Bures-metric-interpretation}
}
reproduces the metric (\ref{g^vac}).

With the simplest choice (\ref{simplest choice}),
it is known that the bulk spacetime becomes AdS with its radius $\ell^2 \Delta$ \cite{Aoki:2017bru}:
\eqn{
g^\tx{vac}_{MN}(X) := \frac{\ell^2\Delta}{z^2} \delta_{MN} ~.
}
Even for general choices of the kernel \eqref{general choice} corresponding to the general flow \eqref{flow_general},
$g^\tx{vac}_{MN}$ also describes the AdS spacetime simply by the symmetry argument.\footnote{One can generalize this statement further: if the flow equation is invariant under the scale transformation such that $\eta\to \lambda^2\eta$ and $x^\mu \to \lambda x^\mu$ as well as the Poincare transformation, the VEV of the metric operator describes the AdS spacetime. }
Such generic choices including the simplest one, however,   give rise to problems for excited states.
For example, it does not reproduce a form of the bulk-to-boundary scalar propagator expected from the AdS/CFT.
Therefore a special choice of the flow equation has been proposed in Ref.~\cite{Aoki:2022lye} to improve properties of the bulk reconstruction, as explained in the following sections.
In this paper, we call such a special flow a conformal flow.

\subsection{Conformal flow and GKP-Witten relation \label{Sec-Special flow}}
Following \cite{Aoki:2022lye}, we take $\nu =\bar\Delta$ together with a normalization of the $d+1$-th coordinate $z$ as $4\eta=\alpha z^2$.
With this special form of the flow equation, 
the radius of the resulting AdS spacetime turns out to be
\eqn{
L^2(d, \Delta ) := \ell^2 \times \frac{\Delta (d-\Delta)}{d+1} ~, \label{L^2(d,Delta)} 
}
and more importantly,
the conformal transformation of the scalar field operator 
\eqn{\delta^\tx{conf}\hat{\vp}^a(x) := -\delta x^\mu \p_\mu \hat{\vp}^a(x) - \frac{\Delta}{d}\qty(\p_\mu \delta x^\mu) \hat{\vp}^a(x)}
with $\delta x^\mu = a^\mu + \omega^{\mu}_{~\nu}x^\nu +\lambda x^\mu + b^\mu x^2 -2 x^\mu (b\cd x)$
becomes a diffeomorphism
transformation of the bulk field operator 
\eqn{\delta^\tx{conf} \hat{\sigma}_0^a(X) = -\bar{\delta}x^\mu \p_\mu \hat{\sigma}_0^a(X) - \bar{\delta}z \p_z \hat{\sigma}_0^a(X)}
with $\bar{\delta}x^\mu =  \delta x^\mu +z^2 b^\mu $, $\bar{\delta} z = (\lambda -2 b\cd x)z$,
which is nothing but the isometry of the AdS spacetime.
A combination of both conformal symmetry at the boundary and AdS isometry in the bulk strongly constraints forms of boundary-to-bulk propagators including all quantum corrections. 
We therefore call this special flow the conformal-AdS flow, or the conformal flow for short. 

For example, the form of the bulk-to-boundary propagator for an $O(N)$ singlet boundary scalar operator $\hat{S}$ can be fixed by the bulk AdS isometry and the boundary conformal symmetry up to an overall constant $C_S$ as 
\eqn{
\La \hat{\sigma}_0^2(X) \hat{S}(y) \Ra_0 = C_{S} \qty(\frac{z}{\abs{x-y}^2 + z^2})^{\Delta_S}
\stackrel{z\to 0}{\simeq} 
\left\{
\begin{array}{ll}
\tilde{C}_S z^{d-\Delta_S}\delta^{(d)}(x-y)~, & \abs{x-y}=0 \\
\\
C_S z^{\Delta_S}  \abs{x-y}^{- 2\Delta_S} ~,& \abs{x-y}\not=0 \\
\end{array}\right. ~, \label{bulk-to-boundary-propagator}
}
where  $\Delta_S >d/2$ is a conformal dimension of $\hat{S}$, 
and  
$\tilde{C}_S=C_S\pi^{d/2}\Gamma \qty(\Delta_S -d/2)/\Gamma \qty(\Delta_S)$.

Let us define the VEV of the scalar operator $\hat{\sigma}_0^2(X)$ in the presence of a small source term $J$ coupled to the scalar operator $\hat{S}$ at the boundary as
\eqn{
\chi_{J} (X) := \La 0\vert \hat\sigma_0^2(X) \exp[\int d^dy\, J(y) \hat{S}(y)]\vert 0\Ra -1
\simeq \int \dd^d y J(y) \sum_{a=1}^N \La \hat{\sigma}_0^a(X)^2  \hat{S}(y) \Ra_0 ~,
}
where a trivial constant is subtracted.
It is easy to check that $\chi_J(X)$ satisfies the massive Klein-Gordon equation in the bulk AdS with the radius (\ref{L^2(d,Delta)}), whose mass is given by $m^2=\Delta_S(\Delta_S-d) / L^2(d,\Delta)$.
For small $z$, we then reproduce the GKP-Witten relation \cite{Gubser:1998bc,Witten:1998qj} as
\eqn{
\lim_{z\to 0} \chi_{J} (X) = z^{d-\Delta_S} \qty[ \tilde{C}_{S} J(x) + \order{z^2} ] + z^{\Delta_S} \qty[  A_{0,J}(x)   + \order{z^2}] ~, \label{GKPW}
}
where 
\eqn{
A_{0,J}(x) :=  C_{S} \int \dd^d y   \frac{J(y)}{\abs{x-y}^{2 \Delta_S}}  
\propto  \int \dd^d y J(y) \La \hat{S}(x)  \hat{S}(y) \Ra_0 \sim \La \hat{S}(x) e^{\int \dd^d y J(y) \hat{S}(y) } \Ra_0  \label{1pt-function}
}
can be interpreted as the expectation value of the operator $\hat{S}$ in the presence of the source term  \cite{Balasubramanian:1998de,Klebanov:1999tb}, since the one-point function vanishes for the conformal symmetry kept unbroken with the vacuum state $\ket{0}$. 

Let us assume the relation (\ref{GKPW}) applies to a generic boundary state $\vert B\rangle$  with $A_J (x) \propto \La B\vert \hat{S}(x) e^{\int \dd^d y J(y) \hat{S}(y) }\vert B \Ra$ replacing $A_{0,J}$.
It tells us that, with $J=0$,
the expectation value of the boundary operator  $\hat{S}$ with conformal dimension $\Delta_S$ generates the correspondent bulk excitation $\chi_B$
as
\eqn{
\chi_B (X):=\La B\vert \hat{\Phi}_S(X)\vert B\rangle \propto z^{\Delta_S} \La B\vert \hat{S}(x)\vert B \Ra ~ \label{bulk-excitation}
}
with some subleading terms that are not explicitly shown, where $\hat{\Phi}_S(X)$ is the corresponding scalar operator in the bulk.
This corresponds to the relation (\ref{BDHM}) implied by the standard bulk reconstruction \cite{Banks:1998dd,Hamilton:2005ju,Hamilton:2006az}.
We take this relation as a guide to identify a bulk theory dual to the boundary theory in the following sections.

For the conformal flow,
the two-point function (\ref{def-G_0}) can be written in terms of a hypergeometric function\cite{Aoki:2022lye}:
\eqn{
G_0(X,X') =~ _2F_1 \qty(\frac{\Delta}{2},\frac{d-\Delta}{2};\frac{d+1}{2};1-U_0^2) ,
\quad
U_0 = \frac{\abs{x-x'}^2 + z^2 + z'^2}{2 z z'}~,
\label{G02F1}
}
where the AdS isometry implies that $G_0 (X,X')$ must be a function of the $SO(1,d+1)$ invariant ratio $U_0$.
In the appendix \ref{App-Hypergeometric function},
the normalization factor in (\ref{def-sigma_0}) is evaluated as
\eqn{
\Upsilon (z) :=  \La \hat{\phi}^2(x;\eta) \Ra_0 = N C_1 z^{-2 \Delta} ~,
\quad
C_1:=C_0  \frac{\Gamma \qty(d-\Delta) \Gamma \qty(d/2)}{\Gamma \qty(\bar\Delta)  \Gamma \qty(d)} .
\label{Upsilon}
}

We may also determine a geometry corresponding to an excited state rather than the vacuum.
For a primary scalar state $\vert S \Ra$,\footnote{This state is generated by the primary scalar operator as $\vert S\Ra = \lim_{\abs{x}\to 0} \hat{S}(x)\vert 0\Ra$. }
for example, we define an operator normalized for this state
as
\eqn{
\hat{\sigma}^a_S(X):= {\hat{\phi}^a(x;\eta)\over \sqrt{\La S\vert \hat{\phi}^2(x;\eta)\vert S\Ra}}~.
\label{normalization_scalar}
}
The corresponding metric operator is defined by
\eqn{\hat{g}^S_{MN}(X):=\ell^2 \sum_{N=1}^N {\partial \hat{\sigma}^a_S(X)\over \partial X^M}
{\partial \hat{\sigma}^a_S(X)\over \partial X^N}~,
\label{metric_scalar}
}
and its expectation value is given by $\La S\vert \hat{g}^S_{MN}(X)\vert S\Ra$,
which can also be interpreted as the Bures information metric.

As an more complicated case, we may consider a mixed state such as the thermal state, which is a target in this paper. In the remaining of this paper,
we define a bulk metric operator for the thermal state which can be interpreted as the Bures metric,
and calculate its thermal average to determine the structure of the spacetime and the corresponding bulk theory.

\section{Bulk geometry from a thermal boundary state \label{Sec-Bulk geometry from thermal boundary state}}
Hereafter we assume that $\vp^a$ is a free massless scalar field, whose action is given by
\eqn{
S_E[\vp]= \int \dd^d x\, \sum_{a=1}^N{\delta^{\mu\nu}\over 2} \p_\mu \vp^a(x) \p_\nu \vp^a(x),
\label{free_action}
}
and calculate the metric $g^{T}_{MN}(X)$ for the thermal state with a temperature $T$ to investigate the corresponding bulk spacetime.
While the conformal dimension is explicitly given by $\Delta=(d-2)/2$ in this case, we keep using $\Delta$ in our formula
instead of $(d-2)/2$ for notational simplicity.

Previously, in Ref.~\cite{Aoki:2020ztd}, the metric $g^T_{MN}(X)$ has been evaluated 
by the flow method and it has been concluded that the metric describes an asymptotically AdS black brane with some unknown matter contribution. However this previous study is unsatisfactory due to the following reasons. First of all, the flow used in Ref.~\cite{Aoki:2020ztd} is the simple Gaussian flow \eqref{simplest choice}, which fails to map the conformal symmetry at the boundary to the AdS isometry in the bulk. Secondly, the thermal average of the metric operator can not be regarded as the information metric due to the inadequate normalization. Both problems seem to make it difficult to draw a clear interpretation on a structure of the bulk theory.

In this paper, we employ the conformal flow with a state dependent normalization suitable for the information metric to calculate the thermal average of the metric operator, $g^T_{MN}(X)$, whose
UV and IR behaviors allows us more explicit interpretation in terms of the AdS/CFT correspondence, as will be seen. 

\subsection{Bulk metric for a thermal state \label{Thermal metric}}
Since thermal correlation functions have $1/T$ periodicity in the Euclidean time direction, the two-point function for a {\it free} theory can be written in terms of the vacuum one (\ref{vacuum-two-point-func.}) as 
\eqn{
\La \hat{\vp}^a (x)\hat{\vp}^b (x') \Ra_T := Z_T^{-1} \tx{Tr} \qty[ e^{-\hat{H}/T} \hat{\vp}^a (x)\hat{\vp}^b (x') ] 
= \sum_{n=-\infty}^{\infty} \La \hat{\vp}^a (\tau + n/T , \vb*{x})\hat{\vp}^b (\tau' , \vb*{x}') \Ra_0 ~,\label{prop_thermal}
}
where $Z_T =\tx{Tr} [ e^{-\hat{H}/T} ]$ and $\hat{H}$ is the Hamiltonian operator associated with the action (\ref{free_action}).\footnote{Eq.~\eqref{prop_thermal} can be justified as follows. 
A $1/T$ periodic function is defined from a non-periodic function $f(x)=\int dp \tilde f(p) e^{i px}$ as
\eqn{f_T(x)= T\sum_{n=-\infty}^\infty \tilde f(p_n) e^{i p_n x}~, \quad p_n =2\pi n T~, }
where $\tilde f(p)$ is the Fourier transformation of $f(x)$. 
While we can construct another $1/T$ periodic function as $g_T(x):=\sum_{n=-\infty}^\infty f(x+n/T)$, the Fourier expansion of $g_T(x)$ implies $g_T(x)=f_T(x)$.
}
Modifying the denominator of (\ref{def-sigma_0}), we have a normalized flowed field operator for the thermal state as\footnote{We do not put a subscript $T$ to $\hat{\sigma}$, since we can easily distinguish it from the vacuum one $\hat\sigma_0$.}
\eqn{
\hat{\sigma}^a(X) :=  \frac{\hat{\phi}^a(x;\eta)}{\sqrt{ \La \hat{\phi}^2(x;\eta) \Ra_T}} = \frac{\hat{\sigma}_0^a(X)}{\sqrt{ \La \hat{\sigma}_0^2(X) \Ra_T}}  ~,
\label{def-sigma}
}
whose thermal expectation value is normalized as
\eqn{
\La \hat{\sigma}^a(X)\hat{\sigma}^b(X)\Ra_T = {\delta^{ab}\over N}, \quad
\La \hat{\sigma}^2(X)\Ra_T = 1.} 
As already mentioned, this normalization is different from the one adopted in \cite{Aoki:2020ztd}.

Using $\hat{\sigma}^a(X)$, the bulk metric is constructed in the same manner as in (\ref{g^vac}) and (\ref{ghat^vac}):
\eqn{
\hat{g}^T_{MN} (X) := \ell^2 \times \sum_{a=1}^N \frac{\p \hat{\sigma}^a(X)}{\p X^M} \frac{\p \hat{\sigma}^a(X)}{\p X^N} ~, ~~~ g^T_{MN}(X) := \La \hat{g}_{MN}^T (X) \Ra_T ~. \label{g^T}
}
Note that the metric operator $\hat{g}^T_{MN}$ at $T\not=0$ is different from $\hat{g}_{MN}$ in (\ref{ghat^vac}), but this $T$ dependent definition is indeed necessary for $\hat{g}^T_{MN}$ to
be interpreted as the Bures information metric as explained below.
Instead of the vacuum state $\ket{0}$ in (\ref{vacuum}), we take the thermofield double state
\eqn{\ket{\tx{TFD}} =  Z_T^{-1/2}\sum_{n} e^{-E_n /2T} \ket{E_n} \otimes \widetilde{\ket{E_n}} ~, ~~~ Z_T = \sum_i e^{-E_n /T}
}
where $\ket{E_n}$'s are energy eigenstates defined by $\hat{H} \ket{E_n} = E_n \ket{E_n}$ and $\widetilde{\ket{E_n}}$'s are their copies. 
With the identity operator  $\tilde{1}$ on the Hilbert space spanned by $\widetilde{\ket{E_n}}$'s, we consider a density operator
\eqn{
\rho_T(X):=& \sum_{a=1}^N \qty(\hat{\sigma}^a(X) \otimes \tilde{1})\ket{\tx{TFD}}\bra{\tx{TFD}} \qty( \hat{\sigma}^a(X)  \otimes \tilde{1}) ~,
}
which satisfies a correct normalization as
\eqn{\Tr[\rho_T(X)]=Z_T^{-1}\sum_n e^{-E_n/T}\La E_n \vert \hat{\sigma}^2(X)\vert E_n \Ra
=\La \hat{\sigma}^2(X) \Ra_T = 1.
}
Since $\rho_T(X) \rho_T(X) = \rho_T(X) /N$ leads to
$\hat{G}= N \dd \rho_T(X) = N \dd X^M \p_M \rho_T(X)$, the bulk metric 
$g_{MN}^T(X)$ in (\ref{g^T}) is reproduced in the same manner as (\ref{Bures-metric-interpretation}).\footnote{If we start with the thermal state without the purification, the corresponding information metric turns out to be more complicated than (\ref{g^T}).}

\subsection{Computing bulk metric \label{Sec-Computing bulk metric}}
The thermal expectation value of the two-point function of the flowed field (\ref{def-sigma_0}) normalized in the vacuum is given by
\eqn{
G(X,X'):= \delta_{ab} \La \hat{\sigma}_0^a(x;\eta) \hat{\sigma}_0^b(x';\eta') \Ra_T = \sum_{n=-\infty}^\infty ~ _2F_1 \qty(\frac{\Delta}{2},\frac{d-\Delta}{2};\frac{d+1}{2};1-U_n^2) \label{two-point}
}
where
\eqn{
U_n = \frac{(\tau -\tau' + n/T)^2 +\abs{\vb*{x}-\vb*{x}'}^2 + z^2 + z'^2}{2 z z'} ~.
}
We define two functions as
\eqn{
\mf{g}(\xi^2):= G(X,X) = \sum_{n=-\infty}^{\infty}   \!\!\! ~_2F_1 \qty(\frac{\Delta}{2},\frac{d-\Delta}{2};\frac{d+1}{2};1-u_n^2) ~, \label{mfg}
}
\eqn{
\mf{f}(\xi^2):=  \sum_{n=-\infty}^{\infty}   \!\!\! ~_2F_1 \qty(\frac{\Delta+1}{2},\frac{d-\Delta +1}{2};\frac{d+3}{2};1-u_n^2)  \label{mff}
}
with $u_n :=1+ n^2 / (2\xi^2)$,
where $\xi:= T z$, which represents a dimensionless coordinate in the bulk direction.

As derived in Appendix \ref{App-Derivatives},
nonzero components of the bulk metric, defined in (\ref{g^T}) can be written in terms of $\mf{g}$ and $\mf{f}$ and their derivative with respect to $\xi^2$ such as $\mf{g}' := \p \mf{g} /\p \xi^2$:
\eqn{
 g_{ii}^T(z) = \ell^2  \eval{ \frac{\p}{\p x^i} \frac{\p}{\p x'^i} \frac{G(X,X')}{ \sqrt{G(X,X)}\sqrt{G(X',X')} } }_{X'\to X} 
 = \frac{L_{d+1}^2}{z^2} \times    \frac{ \mf{f}  }{\mf{g}}  ~, \label{ii}
}
\eqn{
 g_{00}^T(z) - g_{ii}^T(z) &= \ell^2  \eval{ \qty(\frac{\p}{\p \tau} \frac{\p}{\p \tau'} - \frac{\p}{\p x^i} \frac{\p}{\p x'^i} )  \frac{G(X,X')}{ \sqrt{G(X,X)}\sqrt{G(X',X')} } }_{X'\to X} \\
 &= \frac{L_{d+1}^2}{z^2} \times  (-2) \xi^2  \frac{ \mf{f}'  }{\mf{g}}   ~, \label{00-ii}
}
\eqn{
 g_{zz}^T(z) - g_{ii}^T(z)  &= \ell^2  \eval{ \qty(\frac{\p}{\p z} \frac{\p}{\p z'} - \frac{\p}{\p x^i} \frac{\p}{\p x'^i} )   \frac{G(X,X')}{ \sqrt{G(X,X)}\sqrt{G(X',X')} } }_{X'\to X} \\
 &= \frac{L_{d+1}^2}{z^2} \times  \frac{\ell^2}{L_{d+1}^2} \frac{\xi^2}{\mf{g}}  \qty[ \xi^2 \mf{g}'' + \mf{g}'  -\xi^2 \frac{ ( \mf{g}' )^2}{\mf{g}} ] ~. \label{zz-ii}
}
where
\eqn{
L^2_{d+1} := L^2(d, \Delta ) 
= \ell^2 \times \frac{d^2 - 4}{4(d+1)} ~. \label{L^2_{d+1}} 
}

\begin{figure}[tbh]
\begin{center}
\includegraphics[width=16cm]{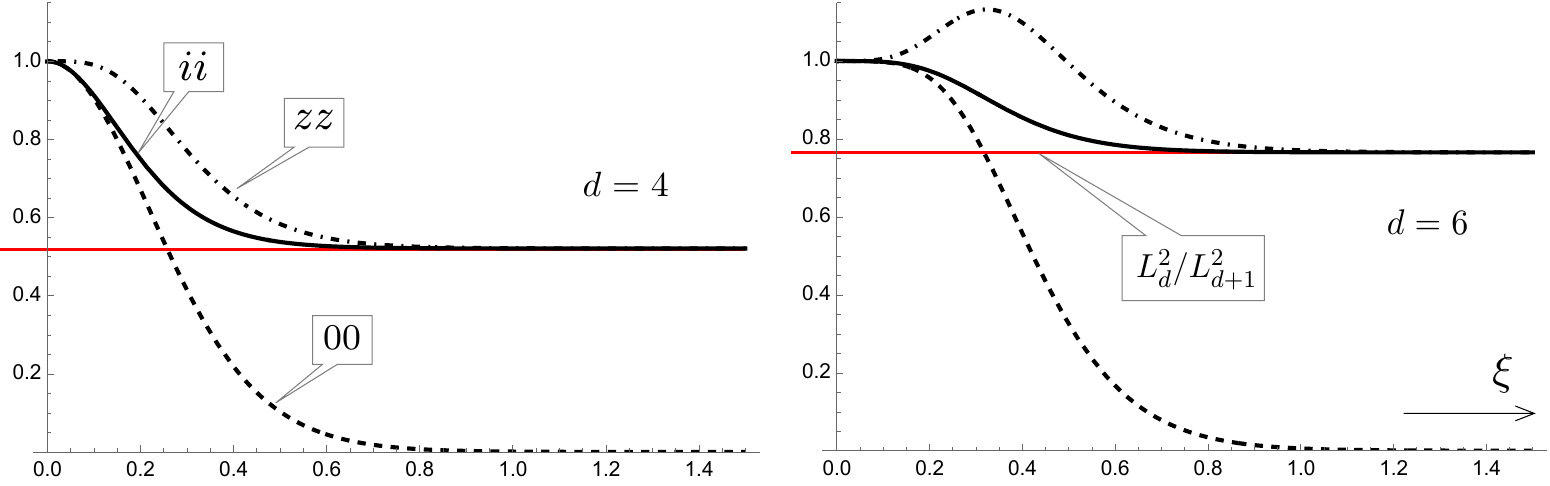}
\caption{Nonzero components of the bulk metric, $g_{ii}$ (solid), $g_{00}$ (dashed) and $g_{zz}$ (dash-dotted), each divided by $L^2_{d+1}/z^2$, as functions of $\xi = T z$ with $d = 4$ (left panel) and $d=6$ (right panel).
The horizontal line depicts $L^2_{d}/L^2_{d+1}= (d+1)^2 (d-3)/ d (d^2 -4)$ for each case. 
For $d = 2(j+1)$ with integer $j\geq 3$, we obtain similar behaviors as in the case at $d=6$. See Fig.\ref{Fig.metric-App} in Appendix~\ref{App-Numerical}.
}
\label{Fig.matric}
\end{center}
\end{figure}
In Fig.~\ref{Fig.matric}, nonzero components of the bulk metric are depicted at $d=4$ and $d=6$.
For even $d$, the infinite summations in (\ref{mfg}) and (\ref{mff}) can be analytically  performed,
as shown in Appendix~\ref{App-Numerical}.
In the UV limit that $z\ll 1/T$, the metric approaches the AdS one with the radius (\ref{L^2_{d+1}}).
In the IR limit that $z \gg 1/T$, on the other hands, $g_{ii}^T$ and $g_{zz}^T$ converges to  $L_d^2 / z^2$ where $L^2_d = L^2(d-1, \Delta -1/2)$ as defined in (\ref{L^2_d}),  while $g_{00}^T$ vanishes exponentially.

In the following, we analytically investigate UV and IR asymptotic behaviors for both even and odd $d$.

\subsection{UV and IR limits \label{Sec-UV and IR limits}}
 We analytically derive UV and IR behaviors shown in Fig.\ref{Fig.matric}.
In particular, we argue that  
the spacetime is conformally equivalent to $\tx{AdS}_2 \times \tx{R}_{d-1}$ in the IR limit.

\subsubsection{UV region $(z \ll 1/T)$}
Using an expansion of the hypergeometric function for small values of $v= 2 \xi^2 /n^2$,
\eqn{
_2 F_1 \qty(a_1,a_1+1;b_1,-\frac{1+2 v}{ v^2} ) =  \frac{\Gamma(b_1) }{\Gamma(b_1 -a_1)\Gamma(1+a_1)} \times v^{2 a_1}   \qty[  1 - 2 a_1 v +  \order{v^2 \ln{v}}] ~, \label{Hypergeometric-expanded}
}
\eqref{mfg} is expanded as $\mf{g} = 1 + \delta \mf{g} $, where
\eqn{
\delta \mf{g} 
=   \frac{4} { \tx{B}\qty( \Delta, d-\Delta) } \frac{\xi^{d-2}}{d-2} \qty{  \zeta (d-2) - (d-2) \zeta(d) \xi^2 + \cl{O}\qty(\xi^4) } \label{mfg-expanded}
}
with $\tx{B}(a,b)$ and $\zeta (s) = \sum_{n=1}^{\infty} n^{-s}$ being the beta function and the zeta function, respectively.
Note that $\zeta (d-2)$ is divergent at $d=3$, which corresponds to the IR divergence of the massless scalar two-point function $1+\delta \mf{g}|_{z\rightarrow 0} \propto \langle \hat{\varphi}^2(x) \rangle_T^{}$ at finite temperature in $d=3$.
By replacing $d$ with $d+2$ and $\Delta$ by $\Delta +1$, we obtain
$\mf{f} =  1 + \delta \mf{f}$, where
\eqn{
\delta \mf{f}  :=&   \frac{4} { \tx{B}\qty( \Delta + 1, d-\Delta +1) } \frac{\xi^{d}}{d}  \qty{  \zeta (d) - d \zeta(d+2) \xi^2  + \cl{O}\qty(\xi^4) } ~.
\label{mff-expanded}
}
Plugging these expressions into (\ref{ii}), (\ref{00-ii}) and (\ref{zz-ii}),
we find the spacetime metric expanded with respect to $\xi=Tz$ as\footnote{
Rigorously speaking, there exist logarithmic terms such as $\xi^{A(d+2 B)} (\ln \xi)^{p}$ with integers $A, B \ge 1$ and $p\ge 1$ coming from terms which are not explicitly shown in the expansion (\ref{Hypergeometric-expanded}). 
We simply omit such terms in \eqref{g-expanded} because they do not appear in the following analyses at the leading and next leading orders.
}
\eqn{g_{MN}^T(z,T) = \bar{g}_{MN}(z) + \delta g_{MN}(z,T) ~, ~~~ \bar{g}_{MN}(z) =  \frac{L_{d+1}^2}{z^2} ~, \label{gbar+deltag} }
\eqn{
\delta g_{MN}(z,T) = \sum_{A, B} \delta g_{MN}^{(A,B)}(z,T) = \frac{L^2_{d+1}}{z^2}  \sum_{A, B} \xi^{A (d + 2 B)} \check{g}_{MN}^{(A,B)} ~ \label{g-expanded}
}
with $A\geq 1$ and $B \geq -1$.
In the UV limit of $z\to 0$, only $\bar{g}_{MN}$ survives, and hence, we get AdS spacetime with the radius (\ref{L^2_{d+1}}).

In this paper, we focus on terms with $A=1$, which have largest corrections to the AdS metric in the UV region.
Their coefficients, $\check{g}^{(1,B)}$ with $B\geq -1$, can be read off from linearized versions of (\ref{ii}), (\ref{00-ii}) and (\ref{zz-ii}) as
\eqn{
\sum_{B} \xi^{d + 2 B} \check{g}_{ii}^{(1,B)} = - \delta \mf{g} + \delta \mf{f} ~, 
\quad
\sum_{B} \xi^{d + 2 B} \qty( \check{g}_{00}^{(1,B)} - \check{g}_{ii}^{(1,B)} ) = - 2 \xi^2  \delta \mf{f}' ~,   
\notag
}
\eqn{
\sum_{B} \xi^{d + 2 B} \qty( \check{g}_{zz}^{(1,B)} - \check{g}_{ii}^{(1,B)} ) = \frac{\ell^2}{L_{d+1}^2}   \qty[ \xi^4 \delta \mf{g}'' + \xi^2 \delta \mf{g}'  ] ~. \label{zz-ii_A=1}
}

The largest contribution in UV is determined at $B=-1$ as
\eqn{
\check{g}_{ii}^{(1,-1)} = \check{g}_{00}^{(1,-1)} =  \frac{ -1 }{d-2} \times  \frac{4  \zeta (d-2) } { \tx{B}\qty( \Delta, d-\Delta) }   ~, 
\quad
 \check{g}_{zz}^{(1,-1)} - \check{g}_{ii}^{(1,-1)} = \frac{ d+1}{d+2}  \times   \frac{  4  \zeta (d-2) } { \tx{B}\qty( \Delta, d-\Delta) } ~. \label{gcheck^(1,-1)_zz}
} 
In Sec.\ref{Sec-scalar}, we will see that this part contains a physical scalar propagating mode, which is absent in the pure Einstein gravity.

The next largest contribution
with $B=0$ becomes\footnote{
For $d=4$, the part with $(A,B) = (2,-1)$ has the same falloff behavior $\xi^{4} \times (L_{d+1}^2/z^2)$. In the current analysis, we simply regard $d$ as a parameter and work with the series expansion  in powers of $\xi^{d+2 B}$.
}
\eqn{
\check{g}_{ii}^{(1,0)}=   \qty( 1 +  \frac{ 4(d+1)}{d^2 -4} ) \times  \frac{  4  \zeta (d) } { \tx{B}\qty( \Delta, d-\Delta) } ~, \label{gcheck^(1,0)_ii}
}
\eqn{
\check{g}_{00}^{(1,0)} - \check{g}_{ii}^{(1,0)} =  (-d) \frac{ 4(d+1)}{d^2 -4}  \times  \frac{ 4  \zeta (d) } { \tx{B}\qty( \Delta, d-\Delta) }  ~, \label{gcheck^(1,0)_00}
}
\eqn{
\check{g}_{zz}^{(1,0)} - \check{g}_{ii}^{(1,0)} =  \frac{-d^2}{4} \frac{ 4(d+1)}{d^2 -4} \times \frac{  4  \zeta (d) } { \tx{B}\qty( \Delta, d-\Delta) }  ~, \label{gcheck^(1,0)_zz}
}
which contains a tensor perturbation, as will be seen
in Sec.\ref{Sec-tensor}.

\subsubsection{IR region $(z \gg 1/T )$}
To investigate the IR region ($\xi = z T \gg 1$),
it is convenient to employ an alternative integral representation of the hypergeometric function
derived in Appendix~\ref{App-Hypergeometric function} as
\eqn{
\!~_2 F_1 \qty(\frac{\Delta}{2},\frac{d-\Delta}{2};\frac{d+1}{2}; 1- u_n^2) = \frac{1}{\tx{B}\qty(d-\Delta, \Delta) } 
\int_0^\infty   {\dd w\over w^{\Delta - {d-2 \over 2}}} \,  h_d (n; w, \xi )   ~,
}
where $h_d (n; w, \xi ):= (2 u_n + w +1/w)^{-d/2}$.
With the Poisson resummation formula $\sum_{n=-\infty}^\infty e^{2\pi ink}=\sum_{k'=-\infty}^\infty\delta (k'-k)$, (\ref{mfg}) turns out to be
\eqn{
\mf{g}(\xi^2) =& \frac{1}{\tx{B}\qty(d-\Delta, \Delta) } \sum_{k=-\infty}^{\infty} \int_0^\infty {\dd w\over w^{\Delta - {d-2 \over 2}}} \,  \tilde{h}_d (k; w, \xi ) ~,\label{mfg-Poisson}
}
where
\eqn{
\tilde{h}_d (k; w, \xi ) :=& \int_{-\infty}^{\infty}\dd n ~ e^{i 2 \pi k n} h_d (n; w, \xi ) = \xi \times \frac{\sqrt{\pi}}{ \Gamma \qty(\frac{d}{2})} \frac{(2 \pi |k| \xi )^{d-1}}{2^{\frac{d-3}{2}}} X_k^{-\frac{d-1}{2}} K_\frac{d-1}{2} ( X_k)
}
with $X_k := 2 \pi |k|  \xi (1+w)/\sqrt{w}$.
Each  $k$ contribution is attributed to the mode with Matsubara frequency $\omega_k = 2 \pi k T$ in the boundary scalar theory.

Since $k\ne 0$ modes are exponentially suppressed for large $\xi$ as
\eqn{
K_\frac{d-1}{2} ( X_k ) = e^{-X_k} \sqrt{\frac{\pi}{2 X_k }} \qty[ 1 + \frac{d(d-2)}{8 X_k } + \order{X_k ^{-2}} ] ~. \label{BesselK}
}
only the $k=0$ contribution 
\eqn{
\tilde{h}_d (0; w, \xi ) = \xi \times \frac{\pi^{1/2} \Gamma \qty(\frac{d-1}{2})}{ \Gamma \qty(\frac{d}{2})}  \qty(\frac{\sqrt{w}}{1+w})^{d-1} 
}
survives in (\ref{mfg-Poisson}) in the IR limit as
\eqn{
\mf{g}  &\simeq \xi \times \frac{\pi^{1/2} \Gamma\qty(\frac{d-1}{2})}{ \Gamma\qty(\frac{d}{2})}  \frac{\tx{B}\qty(d-\Delta -1/2, \Delta -1/2) }{\tx{B}\qty(d-\Delta, \Delta )}  =:  \underline{\mf{g}} ~.
}
By making the replacement that $(\Delta,d)\to (\Delta+1,d+2)$,
we find
\eqn{
\mf{f}  &\simeq \xi \times \frac{\pi^{1/2} \Gamma\qty(\frac{d+1}{2})}{ \Gamma\qty(\frac{d+2}{2})}  \frac{\tx{B}\qty(d-\Delta +1/2, \Delta +1/2) }{\tx{B}\qty(d-\Delta+1, \Delta +1)}  
= \underline{\mf{g}} \times \frac{L^2_{d}}{L^2_{d+1}} 
=:  \underline{\mf{f}} ~,
}
where
\eqn{
L^2_d := L^2(d-1, \Delta-1/2)  =\ell^2{(d-3)(d+1)\over 4d}
\label{L^2_d}}
is the AdS radius (\ref{L^2(d,Delta)}) expected from $(d-1)$-dimensional free scalar CFT, whose conformal dimension is $\Delta- 1/2 = (d-3)/ 2$.
Therefore, the bulk metric in the $\xi = T z \to \infty$ limit becomes
\eqn{
g_{00} \approx 0 ~,~~~
g_{zz}  \approx g_{ii} \approx 
\frac{L^2_{d}}{z^2}   ~. \label{near-horizon-limit}
}
This behavior indicates that the dimensional reduction in the bulk takes place because, in the IR limit corresponding to $z\to \infty$, all the modes with nonzero Matsubara frequency become irrelevant and only the ($d-1$)-dimensional degrees of freedom associated with zero Matsubara frequency are left to form the $d$-dimensional constant $\tau$ hyper-surface whose induced metric is the same as the $d$-dimensional AdS metric with the radius (\ref{L^2_d}).
In other word, the dimensional reduction at the boundary correctly induces the dimensional reduction in the bulk expected by the AdS/CFT correspondence through the conformal flow. 

In the previous work \cite{Aoki:2020ztd}, 
a similar behavior $g_{00} \to 0$ has been found and the similarity to the AdS black brane discussed.
In the following, we propose a different interpretation.
For this purpose, let us take into account the first non-zero Matsubara mode with $|k|=1$.
Since  $X_k$
becomes large as $\xi\to\infty$ and takes a minimum at $w=1$ as
\eqn{X_k = 2 \pi |k| \xi \qty[2 + \frac{(w-1)^2}{4} +\order{(w-1)^3}] ~,}
the integration (\ref{mfg-Poisson}) can be evaluated by the saddle point approximation  around $w = 1$
to result in
\eqn{
\mf{g} - \underline{\mf{g}} 
& \approx   \frac{2}{\tx{B}\qty(d-\Delta, \Delta) }  \frac{\sqrt{\pi}}{\Gamma\qty(\frac{d}{2})}  \qty(\frac{\pi \xi}{2})^{\frac{d-1}{2}}  e^{-4 \pi \xi} =: \mf{g}_\tx{D} 
}
The replacement $(\Delta , d) \to (\Delta + 1, d+2)$ also leads to
\eqn{
\mf{f}_\tx{D} := \mf{f} - \underline{\mf{f}}=\frac{2}{\tx{B}\qty(d-\Delta+1, \Delta+1) } \frac{\sqrt{\pi}}{\Gamma\qty(\frac{d+2}{2})}  \qty(\frac{\pi \xi}{2})^{\frac{d+1}{2}}  e^{-4 \pi \xi} = \frac{\ell^2}{L^2_{d+1}} \times  (\pi \xi) \times \mf{g}_\tx{D} ~.
}
Since $\p_{\xi^2} = (1/2 \xi) \p_\xi$, the leading contribution to $g_{00}$
turns out to be
\eqn{
g_{00}  &\approx  -2 L^2_{d+1} T^2  \times  \frac{-4 \pi}{2 \xi}\frac{\mf{f}_\tx{D}}{\underline{\mf{g}}} 
 = \ell^2 T^2 4 \pi^2  \frac{\mf{g}_\tx{D}}{\underline{\mf{g}}} 
= \frac{L_d^2}{z^2} \times \qty[ 2 \pi T  z_T (\xi) ]^{-2} ~,
}
where
\eqn{
  z_T (\xi) := \frac{1}{2 \pi T}\qty[ \frac{\tx{B}\qty(d-\Delta +\frac{1}{2} ,\Delta+\frac{1}{2})}{2\sqrt{\pi} \Gamma\qty(\frac{d}{2})/\Gamma\qty(d)}  ]^{1/2}
  \frac{e^{+2 \pi \xi} }{\qty(2 \pi \xi )^{\frac{d+1}{4}}} ~.
}
Then, we find
\eqn{
\dd s^2 &\approx \frac{L_d^2}{z^2} \qty[ \frac{ \dd \tau^2 }{(2 \pi T  z_T )^2} + \dd z^2 +  \sum_{i=1}^{d-1}\qty( \dd x^i )^2 ] \\
&=\frac{L_d^2}{z^2} \qty[ \frac{1}{(2 \pi T  z_T )^2} \qty( \dd \tau^2 +   \frac{ \dd z_T^2 }{ \qty(1-\frac{d+1}{8 \pi \xi})^2} ) +  \sum_{i=1}^{d-1}\qty( \dd x^i )^2 ] ~,
}
where $z = \xi /T$ is now regarded as a function of $z_T$.\footnote{Using the Lambert function $W_{-1} \leq -1$, we have
\eqn{
z = -\frac{d+1}{8 \pi T} W_{-1}(v) ~~~\tx{with} ~~~  v:= \frac{-4}{d+1}  \qty[ \frac{\tx{B}\qty(d-\Delta+\frac{1}{2},\Delta+\frac{1}{2})}{2\sqrt{\pi} \Gamma\qty(\frac{d}{2})/\Gamma\qty(d)}  ]^{\frac{2}{d+1}} \qty( 2 \pi T z_T )^{-\frac{4}{d+1}}  ~.
}
}
For $2 \pi \xi \gg (d+1)/4$, it is conformally  equivalent to $\tx{AdS}_2 \times \tx{R}_{d-1}$ as
\eqn{
\dd s^2\approx \dfrac{L_d^2}{z^2} \dd \! \! \stackrel{\circ}{s} \! {}^2, \quad
\dd \! \! \stackrel{\circ}{s} \! {}^2  := \frac{\dd \tau^2 + \dd z_T^2}{z_T^2 /L_T^2} +  \sum_{i=1}^{d-1}\qty( \dd x^i )^2 ~, ~~~~~~ L_T := \frac{1}{2 \pi T} ~. 
\label{IR-limit}
} 
It is well known that  $\tx{AdS}_2 \times \tx{R}_{d-1}$ corresponds to the near-horizon limit of the AdS extremal black brane, whose metric is given by
\eqn{
\dd s^2_\tx{e} = \frac{L^2}{r^2} \qty[ f_\tx{e}(r) \dd \tau^2 + \frac{\dd r^2}{f_\tx{e}(r)} +  \sum_{i=1}^{d-1}\qty( \dd x^i )^2 ]~, 
}
where
\eqn{
f_\tx{e}(r) = 1 -\frac{2(d-1)}{d-2} \qty( \frac{r}{r_\tx{e}} )^d  + \frac{d}{d-2} \qty( \frac{r}{r_\tx{e}} )^{2(d-1)} = d(d-1) \left(1-{r\over r_\tx{e}}\right)^2 h_e\left({r\over r_\tx{e}}\right)
 }
 with $h_e(1)=1$ and $h_e(x^2)>0$.\footnote{
The AdS charged black brane solution of the Einstein-Maxwell theory is given by the factor
\eqn{
 f(r) = 1- \frac{2 \kappa L m }{d-1} \qty(\frac{r}{L})^d + \frac{ \kappa L^2 q^2 }{(d-1)(d-2)} \qty(\frac{r}{L})^{2(d-1)} ~
 }
and the electromagnetic field strength $F = q (r/L)^{d-3} \dd t \wedge \dd r$, where $\kappa$ is the Einstein gravitational constant for the $(d+1)$-dimensional spacetime, $m$ and $q$ are the surface densities of mass and electric charge, respectively.
The extremal solution is obtained when the mass and charge squared are related with each other as
 \eqn{m  = {(d-1)^2 \over (d-2)  \kappa L}  \qty({L\over r_\tx{e}})^d ~, ~~~ q^2 = {d(d-1)\over \kappa L^2 }  \qty({L\over r_\tx{e}})^{2(d-1)}~,}
 where $r_\tx{e}$ is the parameter to be interpreted as the location of the extremal horizon.
 }
Taking the near-horizon limit that $r\simeq r_\tx{e}$ and
making the coordinate transformation as $r_\tx{e}^2/(r_\tx{e}-r) = d (d-1) z_T$, 
we indeed obtain  $\dd s_{\tx{e}}^2\simeq (L^2 / r^2_\tx{e}) \dd \! \! \stackrel{\circ}{s} \! {}^2$
with $L_T^2 = r_\tx{e}^2/d (d-1)$.

\ 

Note that, because of the conformal factor $L_d^2 /z^2 =: \exp{2 \varpi}$, the IR limit has the curvature singularity at $z=\infty$, as in the case of Ref.~\cite{Aoki:2020ztd}.
For instance, the Ricci scalar diverges as\eqn{
R \approx  e^{-2 \varpi} \qty( \stackrel{\circ}{R} -2 d \stackrel{\circ}{\Box} \varpi - d(d-1) \stackrel{\circ}{g} \! {}^{MN} \p_M \varpi \p_N \varpi ) \to \ \stackrel{\circ}{R} \times \frac{z^2}{L_d^2}  ~,
}
where the quantities associated with the metric $\dd \! \! \stackrel{\circ}{s} \! {}^2$ in (\ref{IR-limit}) are denoted  by circles on them and $\stackrel{\circ}{R} \ = -6/L^2_T$ is the scalar curvature   for  AdS$_2$. 

\section{Gravitational theory and bulk-boundary correspondence in UV region \label{Sec-Correspondence between boundary and bulk}}
In the UV region with $z \ll 1/T$, we have the metric perturbation (\ref{g-expanded}) around the AdS spacetime.
In this section, we identify a gravitational theory in which the bulk metric obtained from the boundary theory solves a corresponding equation of motion (EOM), at least, to the leading order perturbations which falloff as $\delta g_{MN} \sim z^{d-4}$ and $ \sim z^{d-2}$.
It turns out that there is a physical scalar mode in the metric, and thus, the pure Einstein gravity cannot be the bulk theory.
Indeed we have found that $f(R)$ gravity does the job, as we will see.
For $f(R)$ gravity, we can take the Einstein frame where the bulk spacetime is to be compared with the AdS black brane geometry.
In addition,
following the relation (\ref{GKPW}), we identify boundary operators corresponding to the bulk physical degrees of freedom.

It should be noted here that $f(R)$ gravity is obtained just as an approximated bulk theory in the UV region.
In fact,  the metric (\ref{IR-limit}) in the IR limit can not be a solution to the EOM of  $f(R)$ gravity.
Therefore, a more general class of modified gravity will be needed to match 
higher order perturbations in the UV region, which are not considered in this paper.

\subsection{Bulk gravitational theory \label{Sec-Bulk theory}}
We are looking for a bulk gravitational theory in the asymptotic AdS region with $\xi=zT \ll 1$,
whose EOM are perturbatively satisfied by the bulk metric determined in the previous section.  
We generally write the EOM of some gravitational theory as  
$\tx{E}_{MN}(g) =0$ and expand the left hand side as
\eqn{
\tx{E}_{MN}(g) = \bar{\tx{E}}_{MN} (\bar{g}) +  \Delta \tx{E}_{MN} (\bar{g}, \delta g)  ~, ~~~  \Delta \tx{E}_{MN} = z^{-2} \sum_{a, b} \xi^{a (d + 2 b )} \check{\tx{E}}_{MN}^{(a,b)} (\bar{g}, \check{g}) ~, \label{EOM-decomposed}
}
where $\bar{g}$ and $\check{g}$ are $\xi$ independent, and $\Delta \tx{E}_{MN}$ contains all contributions from the perturbation characterized by the coefficients $\check{g}_{MN}^{(A,B)}$ with $A\geq 1$ and $B\geq -1$ in (\ref{g-expanded}).
For any given $d\geq 4$, 
in addition to an unperturbed EOM $\bar{\tx{E}}_{MN} (\bar{g},\check{g}) = 0$,  
we consider the perturbation of EOM at the leading order, $\check{\tx{E}}_{MN}^{(1,b)}(\bar{g},\check{g})=0$ with
$b=-1,0$. In this case, we only need linear terms of $\check{g}^{(1,B)}$ with $B=-1,0$ in (\ref{g-expanded}),
which have already been determined in (\ref{gcheck^(1,-1)_zz}), (\ref{gcheck^(1,0)_ii}), (\ref{gcheck^(1,0)_00}) and  (\ref{gcheck^(1,0)_zz}).

In the following,  we show that the metric perturbations from the AdS spacetime for $B=-1$ and $B=0$ are subject to $f(R)$ gravity, whose action is given by
\eqn{
\cl{S} = {1\over 2\kappa} \int \dd^{d+1} x\,  \sqrt{\vert g\vert}  f(R) \label{f(R)-Lagrangian}
}
where $\kappa$ is the $(d+1)$-dimensional gravitational constant and
$g$ is the determinant of the metric.
Here $f(R)$ is a function of the Ricci scalar $R$, which may be assumed to be a polynomial of $R$ as
\eqn{
f(R) =  -2  \Lambda_f + R + \frac{\alpha}{2} R^2 + \cl{O}\qty(R^3) ~, \label{Lambda_in_f(R)}
}
where $\Lambda_f$ and $\alpha$ are constants.
The corresponding EOM become
\eqn{
f g_{MN} = 2 \qty(R_{MN} + g_{MN} \Box -\nabla_M \nabla_N) \p_R f ~.
}
Note that, as seen from the trace of the EOM that
\eqn{
\frac{d+1}{2 d}f =  \qty(  \frac{R}{d} +  \Box ) \p_R f ~,
}
there is a propagating scalar degree of freedom in the theory. 
This can be clearly seen in the Einstein frame, as shown in Appendix \ref{App-f(R) in Einstein frame}.

The unperturbed AdS background $\bar g_{MN}$ must satisfy
\eqn{
\bar{f} \bar{g}_{MN} = 2 \bar{R}_{MN} \overline{\p f}  ~, \label{unperturbed EOM}
}
where $\bar{f} := f(\bar{R})$ and $\overline{\p f} := \p_{R} f(R) |_{R=\bar{R}}$.
Since curvature tensors in the AdS spacetime are obtained from  $\Gamma^A_{z B} = -\delta^A_B/z$ and $\Gamma^z_{\mu\nu}=\delta_{\mu\nu}/z$ as
\eqn{
\bar{R}^{K}_{~MLN} = \frac{2 \bar{R}}{d(d+1)} \delta^K_{[L} \bar{g}_{N]M} ~, ~~~ \bar{R}_{MN} = \frac{\bar{R}}{d+1} \bar{g}_{MN} ~, ~~~ \bar{R} = - \frac{d(d+1)}{L_{d+1}^2} ~, \label{curvatures}
}
(\ref{unperturbed EOM}) leads to
\eqn{
\bar{f}   = \frac{2 \bar{R}}{d+1} \overline{\p f}  ~, \label{To determine Lambda}
}
which is regarded as the equation to determine the cosmological constant  $\Lambda_f$ in (\ref{Lambda_in_f(R)})  for a given $f(R)$ theory.

The perturbation $\delta g_{MN}$ should satisfy
\eqn{
-2 \overline{\p f} \delta G_{MN} + \qty(\bar{f} -\bar{R} \overline{\p f}  ) \delta g_{MN} = 2 \qty(\bar{R}_{MN} + \bar{g}_{MN} \bar{\Box } -\bar{\nabla}_M \bar{\nabla}_N) \overline{\p^2 f} \delta R ~,
}
which, with the aid of (\ref{To determine Lambda}), can be written as
\eqn{
   \delta G_{MN} + \frac{d-1}{d+1} \frac{\bar{R}}{2} \delta g_{MN}  
   =  - \frac{ \overline{\p^2 f}}{\overline{\p f} } \qty(\bar{R}_{MN} + \bar{g}_{MN} \bar{\Box } -\bar{\nabla}_M \bar{\nabla}_N) \delta R  ~. \label{linearized-f(R)}
}
In the Einstein gravity, the right-hand side (r.h.s.) identically vanishes since $\p_R^2 f = 0$ and $(d-1) \bar{R} / 2 (d+1)$ corresponds to the cosmological constant.

\subsubsection{Leading order perturbation $(B=-1)$ \label{Sec-scalar}} 
The leading order metric perturbation with coefficients  in  (\ref{gcheck^(1,-1)_zz}) can be written as
\eqn{
\delta g_{MN}^{(1,-1)} &=  \frac{\bar{g}_{MN} }{d+1} \Phi + L_{d+1}^2 \qty(\bar{\nabla}_M \bar{\nabla}_N - \frac{\bar{g}_{MN}}{d+1} \bar{\Box} ) \Theta_1  
= \frac{\bar{g}_{MN} }{d+1}  \Phi_\tx{ph} + L_{d+1}^2\bar{\nabla}_M \bar{\nabla}_N  \Theta_1  ~, \label{g^(1,-1)}
}
where
\eqn{
\Phi := \bar{g}^{MN} \delta g_{MN}^{(1,-1)} =   \frac{ -4 (d+1) }{d^2-4} \times  \frac{\xi^{d-2} 4  \zeta (d-2) } { \tx{B}\qty( \Delta, d-\Delta) } = - \frac{\ell^2}{L^2_{d+1}} \times  \frac{\xi^{d-2} 4  \zeta (d-2) } { \tx{B}\qty( \Delta, d-\Delta) } ~,   \label{Phi}
}
\eqn{\Theta_1 = \frac{-1}{4(d-1)} \times \Phi ~, 
\quad
\Phi_\tx{ph} := \Phi -   L_{d+1}^2 \bar{\Box} \Theta_1 = \Phi +   2 (d-2)  \Theta_1 = \frac{d}{2(d-1)} \times \Phi ~. \label{scalar}
}
Here (\ref{Box}) is used for $\bar{\Box} \Theta_1$, and
$\Phi_\tx{ph}$ is the only physical perturbation, since
$\bar{\nabla}_M \bar{\nabla}_N  \Theta_1$ in (\ref{g^(1,-1)}) is gauge dependent.\footnote{It can be canceled out by the gauge transformation generated by $\xi^N = - \bar{g}^{NM} \p_M \Theta_1 /2$ and does not contribute the linearized EOM.}
Thus, one can see that there is a physical scalar mode $\Phi_{\rm ph}^{}$ at the leading order of the metric perturbations.

Using results in Appendix \ref{App-Linearized quantities-scalar}, nonzero components of the left-hand side (l.h.s.) of the linearized EOM (\ref{linearized-f(R)}) are evaluated as
\eqn{
 \delta G_{zz} + \frac{d-1}{d+1} \frac{\bar{R}}{2}  \delta g_{zz} = \frac{d}{2} \qty(\delta G_{\sigma \sigma} + \frac{d-1}{d+1} \frac{\bar{R}}{2}  \delta g_{\sigma\sigma} ) =   - \frac{ d (d-1)^2}{2 (d+1)}  \times  \frac{\Phi_\tx{ph}}{z^2} ~, \label{LHS}
}
while nonzero components of the r.h.s. become
\eqn{
  \qty(\bar{R}_{zz} + \bar{g}_{zz} \bar{\Box } -\bar{\nabla}_z \nabla_z) \delta R 
  =  \frac{d}{2}\qty(\bar{R}_\tx{\sigma\sigma} + \bar{g}_{\sigma\sigma} \bar{\Box } -\bar{\nabla}_\sigma \nabla_\sigma ) \delta R 
  = + \frac{ 3 d (d-1)^2}{ (d+1)^2}  \times  \bar{R}\frac{\Phi_\tx{ph}}{z^2} ~.
  \label{RHS}
}
Note that a ratio between $zz$ and $\sigma\sigma$ components is $d/2$ in both \eqref{LHS} and \eqref{RHS}, which also have  the same $z$ and $\xi$ dependence.
Therefore,
the EOM is satisfied, if the second derivative of $f$ at $R=\bar{R}$ is related to the first derivative as
\eqn{
 \overline{\p^2 f} = \frac{d+1}{6}   \frac{\overline{\p f}}{\bar{R}}   ~. \label{second derivative}
}

\subsubsection{Next to leading order perturbation $(B=0)$ \label{Sec-tensor}}
The metric perturbation with the coefficients (\ref{gcheck^(1,0)_ii}), (\ref{gcheck^(1,0)_00}) and  (\ref{gcheck^(1,0)_zz}) is traceless: $\bar{g}^{MN} \delta g_{MN}^{(1,0)} = 0$, and can be written as
\eqn{
\delta g_{MN}^{(1,0)} &= h_{MN}  +  L_{d+1}^2 \qty(\bar{\nabla}_M \bar{\nabla}_N - \frac{\bar{g}_{MN}}{d+1} \bar{\Box} ) \Theta_2  
= h_{MN}  +  L_{d+1}^2 \bar{\nabla}_M \bar{\nabla}_N \Theta_2 ~, \label{g^(1,0)}
}
where
\eqn{
\Theta_2 =  \frac{ -1}{d}  \times \frac{\xi^{d} 4  \zeta (d) } { \tx{B}\qty( \Delta, d-\Delta)  \label{Theta_2}} 
}
is a gauge degree of freedom.
Thus the physical contribution to the perturbation is $h_{MN}$, which satisfies $\bar{g}^{MN} h_{MN} =0$ and $\bar{\nabla}^M h_{MN} = 0$.
Explicitly we have
\eqn{
&h_{ii} = \frac{L^2_{d+1}}{z^2} \times \frac{4 (d+1)}{d^2 - 4} \times \frac{\xi^{d} 4  \zeta (d) } { \tx{B}\qty( \Delta, d-\Delta) }  =\frac{\ell^2}{z^2} \times \frac{\xi^{d} 4  \zeta (d) } { \tx{B}\qty( \Delta, d-\Delta) } ~, \\
& h_{00} = -(d-1) \times h_{ii} ~, ~~~ h_{zz} = 0 ~.
\label{h}
}

As shown in Appendix \ref{App-Linearized quantities-tensor},
this tensor perturbation solves the linearized EOM (\ref{linearized-f(R)}). One can see that
the l.h.s. and the r.h.s. of the linearized EOM (\ref{linearized-f(R)}) vanish separately, which means that $\delta g_{MN}^{(1,0)}$ can be a solution to the pure Einstein gravity. 
Note that, if a scalar mode at this order were presented, the EOM would not be satisfied by (\ref{second derivative}).
 
\subsubsection{Asymptotic equivalence to AdS black brane in Einstein frame \label{Sec-Einstein frame}}
In this subsection, we show that 
the leading deviations $\delta g_{MN}^{(1,B)}$ with $B=-1,0$  from the AdS in the $f(R)$ theory are consistent with those of the AdS black brane at this order in the Einstein frame,
where $f(R)$ theory is transformed to the Einstein gravity plus a massive  scalar field by the Weyl transformation.
Details of calculations in this subsection are given in Appendix \ref{App-f(R) in Einstein frame}.

The metric in the Einstein frame is given by the Weyl transformation as $\tilde{g}_{MN}:= e^{-2s} g_{MN}$, where
$s$ is the scalar field in the Einstein frame.
The leading deviation from the AdS metric, after some calculation, is given by
\eqn{\delta \tilde{g}_{MN}= \tilde h_{MN} +\tilde{L}^2_{d+1} \bar{\tilde{\nabla}}_M \bar{\tilde{\nabla}}_N \tilde{\Theta}, 
\label{delta_gMN_Einstein}}
where $\bar{\tilde{\nabla}}_M$ is the covariant derivative for the AdS metric $\bar{\tilde{g}}_{MN}$ in the Einstein frame with its AdS radius $\tilde L_{d+1}:= e^{-\bar s} L_{d+1}$, defined through $\bar s$, the VEV of $s$.
It is interesting to see that \eqref{delta_gMN_Einstein} contains the tensor perturbation $\tilde h_{MN}:=e^{-2\bar s}h_{MN}$ only without scalar perturbation up to the gauge degree of freedom 
$\tilde{\Theta}:=\Theta_1+\Theta_2$.

Since the scalar excitation around its VEV $\bar s$ behaves as $z^{d-2}$ for small $z$ as explained below (\ref{delta-s}), 
its stress tensor (\ref{EMT in Einstein frame}), which falls off as $\tilde{T}_{MN} \propto z^{2(d-3)}$,
does not contribute to the Einstein equation at the level of perturbation considered in this paper, and thus
we have the vacuum Einstein equation with the cosmological constant (\ref{Einstein frame cosmological constant}).
It is easy to check that
the tensor perturbation $\tilde{h}_{MN}$ is s solution to the linearized vacuum Einstein equation.

We now show that the tensor perturbation $\tilde{h}_{MN}$ is consistent with the leading order deviation of the AdS black brane solution from the pure AdS solution, whose metric is given by
\eqn{ds^2_\tx{bb}={\tilde{L}^2_{d+1}\over z^2}\left[ f(z) d\tau^2 +{dz^2\over f(z)} +\sum_i (dx^i)^2\right]
\label{metric_bb}}
with $f(z) = 1- (z/z_H)^d$, and the associated Hawking temperature is given by $T_H=d/(4\pi z_H)$, so that the leading deviation form the pure AdS is given by
\eqn{\delta g_{00}^\tx{bb}= -{\tilde{L}_{d+1}^2\over z^2}\left({z\over z_H}\right)^d~, \quad
\delta g_{zz}^\tx{bb}= {\tilde{L}_{d+1}^2\over z^2}\left[\left({z\over z_H}\right)^d +O\left(z^{2d} \right)\right] ~, \quad
\delta g_{ii}^\tx{bb} =0.
\label{delta-g_bb}
}
With an identification that
\eqn{T_H = T\times {d\over 4\pi}\left({4d(d+1)\over d^2-4}\times {4\zeta(d)\over B(\Delta, d-\Delta)}\right)^{1/d}~,
\label{Hawking-T}}
we have
\eqn{\tilde{h}_{ii}={\tilde{L}_{d+1}^2\over z^2}\left({z\over z_H}\right)^d\times{1\over d} ~, \quad \tilde{h}_{00}=-(d-1)\tilde{h}_{ii}~, \quad \tilde{h}_{zz} =0 ~,
\label{tilde-h}}
which looks different from \eqref{delta-g_bb}.
It is easy to see, however, that \eqref{delta-g_bb} and \eqref{tilde-h} are equivalent with each other by the gauge transformation as
\eqn{\delta g_{MN}^\tx{bb} =\tilde{h}_{MN} +\bar{\tilde{\nabla}}_M \bar{\tilde{\nabla}}_N \tilde{\Theta}^\tx{bb} ~, \quad
\tilde{\Theta}^\tx{bb}=\left({z\over z_H}\right)^d\times{1\over d^2}~.
\label{h-bb-equivalence}}
Thus the tensor perturbation of the metric generated by the conformal flow in the Einstein frame is the AdS black brane at the first order of the small $z$ expansion.

\subsection{Boundary operators and corresponding bulk degrees of freedom \label{Sec-Boundary operators and correspondent bulk degrees of freedom}}
In this subsection, we will investigate the relation (\ref{bulk-excitation}) between boundary operators and the bulk physical degrees of freedom by determining its proportionality coefficient.
We here define the r.h.s. of (\ref{bulk-excitation})
by regularizing it with the smearing, and then, subtracting its potentially divergent vacuum expectation value as 
\eqn{
  \La \cl{O}(\hat{\vp}(x)) \Ra_T^\tx{finite}  
:=  \qty[  \La \cl{O}(\hat{\phi}(x;\eta)) \Ra_T - \La \cl{O}(\hat{\phi}(x;\eta)) \Ra_0 ]_{\eta \to 0}  \label{finite}
}
for any composite operator $\cl{O}(\hat{\vp}(x))$ in the boundary CFT.

We start with the tensor degrees of freedom obtained in Sec.\ref{Sec-tensor} . We then move to the gravitational scalar degrees of freedom obtained in Sec.\ref{Sec-scalar}, in order to argue  why it cannot be an ordinary non-gravitational scalar field in the Einstein frame.

\subsubsection{Bulk tensor}
The bulk tensor degrees of freedom $z^2 \times h_{MN} \propto z^{d}$ are expected to be related to 
the boundary stress tensor with conformal dimension $d$, which is given by
\eqn{
\hat{T}_{\mu \nu} & = \sum_{a=1}^N \qty[ \p_\mu \hat{\vp}^a \p_\nu \hat{\vp}^a  -\frac{\delta_{\mu \nu}}{2} \delta^{\rho \sigma} \p_\rho \hat{\vp}^a \p_\sigma \hat{\vp}^a ] ~.
}
Since 
\eqn{
\sum_{a=1}^N   \delta^{\rho \sigma} \La \p_\rho \hat{\vp}^a \p_\sigma \hat{\vp}^a \Ra_T^\tx{finite} 
=&\qty[ \Upsilon(z)   T^2 \frac{\tx{B}\qty(d-\Delta +1, \Delta +1)}{\tx{B}\qty(d-\Delta, \Delta)} d \times \qty( \xi^{-2} \delta \mf{f} \times d -2 \mf{f}') ]_{z\to 0} = 0 ~,
}
where we have used (\ref{Gii}), (\ref{dG/dtdt-dG/dxdx}), (\ref{mff-expanded}) and the definition of $\Upsilon (z)$ in (\ref{Upsilon}),
we find
\eqn{
\La \hat{T}_{ii}  \Ra_T^\tx{finite}  
&= 
\qty[ \Upsilon(z)  T^2 \frac{\tx{B}\qty(d-\Delta +1, \Delta +1)}{\tx{B}\qty(d-\Delta, \Delta)} d \times \xi^{-2} \delta \mf{f} ]_{z\to 0}  = \frac{4 \zeta(d) N C_1 T^{d}}{\tx{B}\qty(d-\Delta, \Delta)}  ~,
}
\eqn{
\La \hat{T}_{00}  \Ra_T^\tx{finite}  
&= 
\qty[ \Upsilon(z)  T^2 \frac{\tx{B}\qty(d-\Delta +1, \Delta +1)}{\tx{B}\qty(d-\Delta, \Delta)} d \times \qty( \xi^{-2} \delta \mf{f}  -2 \mf{f}') ]_{z\to 0} = -(d-1) \times \La \hat{T}_{ii}  \Ra_T^\tx{finite} 
}
so that $\delta^{\mu\nu} \La \hat{T}_{\mu \nu}  \Ra_T^\tx{finite} = 0 $.
Comparing the above with (\ref{h}),  
we find 
\eqn{
\frac{z^2}{\ell^2} \times h_{\mu \nu} =  z^d    \times \frac{\La \hat{T}_{\mu \nu}  \Ra_T^\tx{finite} }{NC_1} ~. \label{GKPW-for-tensor}
}

\subsubsection{Bulk scalar}
The bulk scalar degree of freedom $\Phi_\tx{ph} \propto z^{d-2}$ are  related to $\hat{S} := \sum_{a=1}^N  \hat{\vp}^a   \hat{\vp}^a$, 
whose conformal dimension is $2 \Delta = d-2$.
Using   (\ref{mfg-expanded}), we obtain
\eqn{
 \La  \hat{S}  \Ra_T^\tx{finite}  
=&\qty[ \Upsilon(z)  \times  \delta \mf{g}   ]_{z\to 0} = \frac{4 \zeta (d-2)}{\tx{B}(\Delta , d-\Delta)} \frac{N C_1 T^{2 \Delta}}{d-2} ~.
}
Comparing this with (\ref{scalar}), we find the relation that
\eqn{
 \frac{L^2_{d+1}}{\ell^2} \times \Phi_\tx{ph}
= \frac{- d(d-2)}{2(d-1)}  \times z^{2\Delta} \times   \frac{\La  \hat{S}  \Ra_T^\tx{finite}  }{NC_1} ~,\label{GKPW-for-scalar}
}
or equivalently, 
\eqn{
 \frac{L^2_{d+1}}{\ell^2} \times \Phi 
= -(d-2)  \times  z^{2\Delta} \times  \frac{\La  \hat{S}  \Ra_T^\tx{finite}  }{NC_1}   ~.  
}

Note that the operator $\hat{S}$ is the scalar operator with the smallest conformal dimension $2 \Delta$  among scalar operators with nonzero thermal expectation value.
While the operator $\hat{\vp}^a$ has the smaller conformal dimension $\Delta$, its thermal expectation value vanishes due to the unbroken $O(N)$ symmetry.
This is a reason why the simplest possibility, the Einstein gravity with a minimally coupled scalar $\vs$ discussed in Appendix \ref{App-Einstein+scalar}, cannot be the dual bulk theory.
The scalar field needs to behave as $\vs = \vrho \times  z^\Delta$ with $\vrho$ being a constant, 
and thus, the relation (\ref{bulk-excitation}) with $\La \vp^a \Ra_T = 0$ 
leads to $\vrho = 0$, which excludes this simplest possibility.

\section{Conclusions \label{Sec-Summary}}

In this paper, we constructed the $(d+1)$-dimensional bulk Euclidean spacetime from the  massless free scalar theory on the $d$-dimensional flat boundary at finite temperature $T$, by a conformal flow with the smearing size $z$.
It is important to note that the bulk geometry is emergent from the bulk metric operator, whose VEV is interpreted as the Bures information metric associated with the boundary theory.
This bulk reconstruction from  massless free scalar at finite $T$ holds at  $d \geq 4$  due to its IR behaviors.

Applying the conformal flow to the boundary thermal state,
we have computed the bulk information metric whose nonzero components are given in (\ref{ii}), (\ref{00-ii}) and (\ref{zz-ii}).
The resultant bulk metric in the UV region is  perturbatively expanded in (\ref{g-expanded}) with coefficients in  (\ref{gcheck^(1,-1)_zz}), (\ref{gcheck^(1,0)_ii}), (\ref{gcheck^(1,0)_00}) and  (\ref{gcheck^(1,0)_zz}).
As depicted in Fig.\ref{Fig.matric},
the dimensional reduction takes place in the IR region with $g_{00}$ vanishing exponentially and we are left with the $d$-dimensional constant $\tau$ hypersurface whose induced metric  becomes the $d$-dimensional AdS  with the radius $L_d$ in (\ref{L^2_d}),
since only the degrees of freedom with vanishing Matsubara frequencies remain dynamical in the IR limit and are the same as those in $(d-1)$-dimensional massless scalar theory.
With a closer look at the exponential decay of $g_{00}$,
we have found that the IR asymptotic behavior is controlled by the first Matsubara frequency $2 \pi T$ and the spacetime is conformally equivalent to $\tx{AdS}_2 \times \tx{R}_{d-1}$,  which is the near-horizon limit of extremal black brane.

In the UV region,  
the leading order perturbations with coefficients in (\ref{gcheck^(1,-1)_zz}) contain a scalar propagating mode, which cannot be gauged away.
We have shown that this mode can be accounted for by the $f(R)$ gravity, whose linearized EOM is given in  (\ref{linearized-f(R)}), and further, the tensor mode in the perturbation with coefficients (\ref{gcheck^(1,0)_ii}), (\ref{gcheck^(1,0)_00}) and  (\ref{gcheck^(1,0)_zz}) also solves the EOM.
The unknown proportionality coefficient in the relation (\ref{bulk-excitation}) between a bulk excitation and a boundary operator is determined as (\ref{GKPW-for-tensor}) for the tensor mode and as (\ref{GKPW-for-scalar}) for the scalar mode.
Seen in the Einstein frame, the bulk spacetime is asymptotically equivalent to the AdS black brane solution of the vacuum Einstein equation.

\

As mentioned below (\ref{GKPW-for-scalar}), the bulk theory dual to the boundary one cannot be the Einstein gravity coupled to an ordinary scalar field for the relation (\ref{bulk-excitation}) to be satisfied, and thus, the modified gravity is necessary.
It is reasonable to imagine similar phenomena also happen for generic exited stats of the boundary theory. To study geometry corresponding to excited states, a normalization of the smeared field 
is important to define the metric operator whose excited state expectation value can be interpreted as the information metric. For the scalar case, see \eqref{normalization_scalar} and \eqref{metric_scalar}.  
We leave this problem for our future studies, in order to see whether relations (\ref{GKPW-for-tensor}) and (\ref{GKPW-for-scalar}) hold for such generic states.
Having nothing else than the metric field in the bulk seems reasonable in the sense that, 
in our approach with the current model of real scalar field, 
the bulk metric defined by (\ref{g^T}) is the only object that allows the information-theoretical interpretation.

\ 

Lastly, let us emphasize that $f(R)$ gravity is not the exact bulk theory in all regions.
Metric perturbations  that fall off differently from  $z^{d-4}$ or $z^{d-2}$ may no longer solve the EOM of $f(R)$ gravity, and we indeed explicitly confirmed that the metric in the IR limit (\ref{IR-limit}) is not an asymptotic solution of $f(R)$.
Therefore, more general class of modified gravity theories needs to be considered.
In principle, we could find a true bulk  theory by imposing the condition \eqref{EOM-decomposed} order by order.  
It would be interesting to compare such a bulk theory with the higher spin theory, which is conjectured in \cite{Klebanov:2002ja} to be dual to the boundary $O(N)$ scalar theory.

\begin{acknowledgments}
\noindent
J.B. acknowledges support from the International Research Unit of Quantum
Information (QIU) of Kyoto University Research Coordination Alliance,
and also would like to thank the Yukawa Institute for Theoretical Physics
at Kyoto University for support and hospitality by the long term visitor program during the initial stage of this research.
This work has been supported in part by the NKFIH research Grant K134946, by KIAS Individual Grants, Grant No. 090901, and by the JSPS Grant-in-Aid for Scientific Research (No. JP22H00129).
\end{acknowledgments}

\appendix
\section{Normalization and integral representation of the bulk two-point function \label{App-Hypergeometric function}}
We first calculate the normalization factor $\Upsilon (z)$ in \eqref{Upsilon}, which is expressed in the Fourier space as
\eqn{
\Upsilon (z)=\omega\int {\dd^d p\over (2\pi)^d} {1\over (p^2)^{\bar\Delta}}\tilde S^2(p;\eta)
= N C_1 z^{-2\Delta},
}
where
\eqn{
C_1 := {C_0 2^{3-d}\over \Gamma(\bar\Delta) \Gamma(\Delta) \Gamma({d\over 2})}
\int_0^\infty \dd p\, p^{d-1} K_{\bar\Delta}^2(p).
}
Using an integration formula,
\eqn{
\int_0^\infty \dd p \, p^b K_a^2(p) = 2^{b-2}{\Gamma\left(\frac{b+1-2a}{2}\right) \Gamma\left(\frac{b+1+2a}{2}\right) \Gamma^2\left(\frac{b+1}{2}\right)\over \Gamma(b+1)} ~,
}
we obtain 
\eqn{
C_1= C_0{\Gamma(d-\Delta)\Gamma\left({d\over 2}\right)\over \Gamma(\bar \Delta)\Gamma(d)} ~.
}

We next derive an alternative integral representation of the bulk two-point function (\ref{G02F1}).
We first evaluate a bulk-to-boundary two-point function in the Fourier space as 
\eqn{
\La \hat{\sigma}^a_0 (X) \hat{\vp}^b(0) \Ra_0 = {\delta^{ab}\omega\over \sqrt{\Upsilon(z)}}
\int{\dd^d p\over (2\pi)^d}
{e^{i p x}\over (p^2)^{\bar\Delta} }\tilde S(p;\eta) 
=  {\delta^{ab}C_2 \over x^{d-2\over 2}}\int_0^\infty \dd p\, p^{\Delta} J_{d-2\over 2}(xp) K_{\bar\Delta}(zp),
}
where
\eqn{C_2:={C_0 \over \sqrt{NC_1}}{2^{\bar\Delta -\Delta} \Gamma(\bar\Delta)\over \Gamma(\Delta) x^{d-2\over 2}}~, 
\quad
\int_0^\pi \dd \theta\, \sin^{d-2}\theta\,  e^{i px\cos\theta} = {\sqrt{\pi} 2^{d-2\over 2}\Gamma({d-1\over 2})\over (xp)^{d-2\over 2}} J_{d-2\over 2}(x p) ~.
}
Using an integration formula
\eqn{
\int_0^\infty \dd p\, p^{\mu-\nu+1} J_\mu(ap) K_\nu(bp) ={ (2a)^\mu (2b)^{-\nu} \over (a^2+b^2)^{\mu-\nu+1}}\Gamma(\mu-\nu+1) ~,
}
we obtain
\eqn{
\langle \hat\sigma_0^a(X)\hat\varphi^b(y)  \rangle_0 &= {\delta^{ab} C_0 \over\sqrt{N C_1 } }\qty({z \over z^2 +|x-y|^2})^{\Delta} ~,
}

Smearing the remaing $\hat \vp^b(y)$ further, the bulk two-point function (\ref{G02F1}) becomes
\eqn{
G_0(X_1,X_2) &={ 1 \over \sqrt{\Upsilon(z_2) }}\sum_{a=1}^N  \int \dd^d y\, S(x_2-y;\eta_2) \langle  \hat\sigma_0^a(x_1;\eta_1)\hat\varphi^a(y)\rangle_0 \\
&={\Gamma(d)\over \pi^{d/2}\Gamma(d/2)} 
\int \dd^d y\, \qty({ z_1 \over z_1^2+|x_1-y|^2 })^\Delta \qty({ z_2  \over z_2^2+|x_2-y|^2  })^{d-\Delta} .
}
Introducing two Feynman parameters and doing the Gaussian integral with respect to $y$, we obtain
\eqn{
&G_0(X_1,X_2) ={z_1^\Delta z_2^{d-\Delta}\over \tx{B}(\Delta,d-\Delta)\pi^{d/2}\Gamma(d/2)}
\int_0^\infty \dd s\, s^{\Delta-1} \int_0^\infty \dd r\, r^{d-\Delta-1} \int \dd^d y \\
&\times \exp\qty{ -s(z_1^2+|x_1-y |^2)-r (z_2+|x_2-y|^2) }
= {z_1^\Delta z_2^{d-\Delta}\over \tx{B}(\Delta,d-\Delta)\Gamma(d/2)}
\int_0^\infty \dd s\, s^{\Delta-1}\\
&\times   \int_0^\infty \dd r\, r^{d-\Delta-1} (s+r)^{-d/2} \exp[ -s z_1^2 -r z_2^2 -{rs\over s+r}|x_1-x_2|^2].
}
Changing  the integral variables as $s=(1-\beta) t$, $ r=\beta t$ and integrating the overall scale $t$ of the Feynman parameters, we get the final result:
\eqn{
G_0(X_1,X_2) &={(z_2/z_1)^{\bar\Delta} \over \tx{B}(\Delta,d-\Delta)} \int_0^1 \dd \beta {(1-\beta)^{\Delta-1} \beta^{d-\Delta-1}\over \left[ {1-\beta \over z_2/z_1} +\beta z_2/z_1 +\beta(1-\beta) {|x_1-x_2|^2\over z_1 z_2}\right]^{d/2}}\\
&= {1\over \tx{B}(\Delta,d-\Delta)}\int_0^\infty \dd w\, {w^{\bar\Delta-1} \over (2U_0 + w+ 1/w)^{d/2}} ~,
}
where $w =\frac{ (z_2/z_1) \beta}{1-\beta}$, and $U_0$ is the $SO(1,d+1)$ invariant ratio in \eqref{G02F1}.

Comparing the above result with (\ref{G02F1}), we find a new and alternative integral formula for a hypergeometric function in a special case as
\eqn{
{}_2F_1\left({\Delta\over 2}, {d-\Delta\over 2};{d+1\over 2}; 1-U_0^2\right)
= {1\over \tx{B}(\Delta,d-\Delta)}\int_0^\infty \dd w\, {w^{d/2-\Delta-1} \over (2U_0 + w+ 1/w)^{d/2}} 
\label{Integral_rep_2F1}
}
which is valid for arbitrary $d$ and $\Delta$ as long as the integral is convergent,
{\it i.e.} $d > \Delta > 0$.

\section{Derivatives of two-point function \label{App-Derivatives}}
In order to obtain expressions (\ref{ii}), (\ref{00-ii}) and (\ref{zz-ii}) of the bulk metric,
derivatives of the two-point function for the flowed field defined in (\ref{def-sigma}) are rewritten as follows.

Using \eqref{Integral_rep_2F1},
the two-point function (\ref{two-point}) is given by
\eqn{
G (X,X') =\frac{1}{\tx{B}(\Delta, d-\Delta)}  \sum_{n=-\infty}^{\infty}  \int_0^\infty {\dd w\over w^{\Delta - {d-2 \over 2}}} \,  f(d/2 , U_n) ~.
}
where
\eqn{
f(b,U):= (2 U + w +1/w)^{-b} ~, \quad 
\p_U f(b,U) = -2 b \times f(b+1,U).
}
Thus
\eqn{
  \p_{X^M} f( d/2 , U_n) |_{X'\to X} =   - d\,     f(d/2 +1 , u_n) \times \p_{X^M} U_n |_{X'\to X}   ~, \label{DI}
}
\eqn{
 \p_{X^M} \p_{X'^N} f( d/2 , U_n) |_{X'\to X} 
  =&  -d   \bigg[ -(d+2)  f( d/2 +2 , u_n) \times  (\p_{X^M} U_n ) ( \p_{X'^N} U_n ) |_{X'\to X}  \\
  &+  f(  d/2 +1 , u_n) \times \p_{X^M} \p_{X'^N} U_n |_{X'\to X}  \bigg] \label{DDI}
}
where  
\eqn{
 \p_{x^i} U_n |_{X' \to X} = 0 ~, ~~~  \p_{x'^i}\p_{x^j} U_n |_{X' \to X}  =-\delta_{ij} \frac{T^2}{\xi^2} ~,
}
\eqn{
 \p_{\tau} U_n |_{X' \to X} = + \frac{n T}{\xi^2} ~, ~~~ 
 \p_{\tau'} U_n |_{X' \to X}  = -\frac{n T}{\xi^2} ~, ~~~ 
 \p_{\tau'}\p_{\tau} U_n |_{X' \to X}   = -\frac{T^2}{\xi^2} ~,
}
\eqn{
 \p_{z} U_n |_{X' \to X}  = \p_{z'} U_n |_{X' \to X}  = -\frac{n^2 T}{2 \xi^3} ~, ~~~  \p_{z}\p_{z'} U_n |_{X' \to X}  = - \frac{T^2}{\xi^2} \qty(1 - \frac{n^2}{2 \xi^2}) ~.
}
The expressions in (\ref{DI}) and (\ref{DDI}) can be simplified with
\eqn{
-2b\times f( b+1 ,u_n) = \qty( \frac{\p u_n}{\p \xi^2} )^{-1} \frac{\p}{\p \xi^2}  f( b ,u_n)  = - \frac{2 \xi^4}{n^2} \frac{\p}{\p \xi^2}  f(  b ,u_n)
}
for $n\ne 0$,
so that
\eqn{
 \eval{ \frac{\p}{\p z} G(X,X')}_{X'\to X} =   \eval{ \frac{\p}{\p z'} G(X,X')}_{X'\to X} =T   \times \xi  \mf{g}' , \label{dG/dz}
}
\eqn{
 \eval{\frac{\p}{\p x^i} \frac{\p}{\p x'^i} G(X,X')}_{X'\to X} = T^2 \frac{\tx{B}\qty(d-\Delta +1, \Delta +1)}{\tx{B}\qty(d-\Delta, \Delta)} d \times \xi^{-2} \mf{f} ~,
 \label{Gii}
}
\eqn{
 \eval{\qty(\frac{\p}{\p \tau} \frac{\p}{\p \tau'} - \frac{\p}{\p x^i} \frac{\p}{\p x'^i} )G(X,X')}_{X'\to X} = T^2  \frac{\tx{B}\qty(d-\Delta +1, \Delta +1)}{\tx{B}\qty(d-\Delta, \Delta)}  d \times (-2)  \mf{f}' \label{dG/dtdt-dG/dxdx},
}
\eqn{
 \eval{ \qty(\frac{\p}{\p z} \frac{\p}{\p z'} - \frac{\p}{\p x^i} \frac{\p}{\p x'^i} )G(X,X')}_{X'\to X} = T^2  \times \qty( \xi^2 \mf{g}'' + \mf{g}' ) \label{dG/dzdz-dG/dxdx},
}
where we have used a fact that $f(b,u_0)$ is $\xi$-independent, and the $n=0$ contribution exists  only in \eqref{Gii}. Note also that
\eqn{
d\times {\tx{B}(d-\Delta+1,\Delta+1)\over \tx{B}(d-\Delta,\Delta)}={L_{d+1}^2\over \ell^2}.
}

\section{Numerical evaluation of bulk metric \label{App-Numerical}}
\begin{figure}[t]
\begin{center}
\includegraphics[width=16cm]{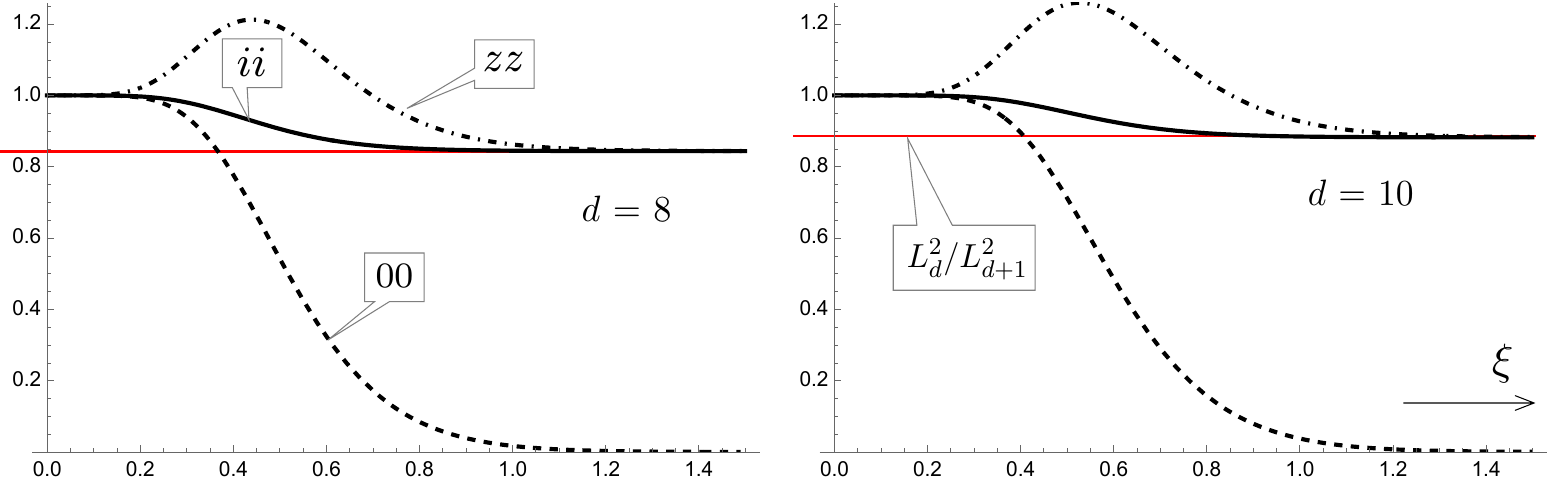}
\caption{Nonzero components of the metric $g_{ii}$ (solid), $g_{00}$ (dashed) and $g_{zz}$ (dash-dotted), each divided by $L^2_{d+1}/z^2$, as functions of $\xi = T z$ with $d = 8$ (left panel) and $d=10$ (right panel).
The horizontal line depicts $L^2_{d}/L^2_{d+1}= (d+1)^2 (d-3)/ d (d^2 -4)$ for each case. 
}
\label{Fig.metric-App}
\end{center}
\end{figure}

By restricting ourselves to the case for an even integer $d= 2 (j+1)$ with $j\geq 1$ integer,
the summation in the bulk metric can be explicitly performed.
Eq.~(\ref{mfg}) can be written as 
\eqn{
\mf{g}=  \frac{(\pi \xi)^{d}}{\tx{B}(\Delta , d-\Delta)} \int_0^\infty {\dd w\over w^{\Delta - {d-2 \over 2}}} \,  S_d \qty(\gamma)  ~, 
\quad
S_d (\gamma) := \sum_{n=-\infty}^{\infty} (\pi^2 n^2 + \gamma^2 )^{-d/2} ~,
\label{S_d in mfg}
}
where $\gamma := \pi \xi \sqrt{2+w+1/w} $.
By a replacement that $(\Delta,d)\to (\Delta+1,d+2)$,
we obtain $\mf{f}$ in (\ref{mff}).
A summation with respect to $n$ can be explicitly performed at $d=2$ as
\eqn{
S_{2}(\gamma) = \sum_{n=-\infty}^{\infty} \frac{\pi^{-2}}{n^2 + (\gamma/\pi)^2} = \frac{1}{\gamma \tanh (\gamma)} ~.
}
Since $\p_{\gamma^2} S_d (\gamma) = -\dfrac{d}{2} \times S_{d+2} (\gamma) $
for $d=2(j+1)$, we find
\eqn{
S_{d}(\gamma) =  \frac{\p_{\gamma^2}^j S_2}{(-1)^j j!} ~. \label{S_d}
}
In Fig.\ref{Fig.matric}, the nonzero components of the bulk metric for $j=1$ and $j=2$ are presented.
As is seen in Fig.\ref{Fig.metric-App}, the metric for $j\geq 3$ has the similar behavior to the $j=2$ case.
\section{Linearized quantities \label{App-Linearized quantities}}
In order to check if the bulk metric from the boundary CFT solves the linearized  EOM of $f(R)$ theory (\ref{linearized-f(R)}), explicit expressions of perturbed quantities like $\delta R$ and $\delta G_{MN}$ are needed.
We here compute them only for the physical parts of the perturbations such as a scalar $\Phi_\tx{ph}$ and a
symmetric, traceless and transverse tensor $h_{MN}$,
since 
parts like $\bar{\nabla}_M \bar{\nabla}_N \Theta_{1,2}$ that can be gauged away automatically satisfy the linearized EOM for the diffeomorphism invariant theory.

\subsection{Scalar $\delta g_{MN} = \bar{g}_{MN} \Phi_\tx{ph} / (d+1)$ \label{App-Linearized quantities-scalar}}
For the scalar perturbation, the Ricci tensor is written as
\eqn{
\delta R_{MN} &= -\frac{1}{2(d+1)} \qty{  (d-1) \bar{\nabla}_M \p_N  + \bar{g}_{MN} \bar{\Box} } \Phi_\tx{ph}  ~.
}
We therefore obtain
\eqn{
\delta R &= - \frac{d}{d+1} \qty( \bar{\Box} - \frac{d+1}{L_{d+1}^2} )  \Phi_\tx{ph} ~,\\
\delta G_{MN} &= \delta R_{MN} - \frac{\bar{g}_{MN}}{2} \qty[  \frac{\bar{R} \Phi_\tx{ph} }{d+1} + \delta R ] 
= -\frac{d-1}{2(d+1)} \qty{ \bar{\nabla}_M \p_N  - \bar{g}_{MN} \bar{\Box} }\Phi_\tx{ph} ~.
}
Since
\eqn{
\bar{\Box} \Phi_\tx{ph} = \frac{X(X-d)}{L^2_{d+1}} \times \Phi_\tx{ph} ~, \label{Box}
}
\eqn{
\qty( \bar{\nabla}_z \bar{\nabla}_z  - \bar{g}_{zz} \bar{\Box} ) \Phi_\tx{ph} &= X d \times \Phi_\tx{ph} /z^2 ~, \\
\qty( \bar{\nabla}_\sigma \bar{\nabla}_\sigma  - \bar{g}_{\sigma \sigma} \bar{\Box} ) \Phi_\tx{ph} &= X(d- X -1) \times \Phi_\tx{ph} /z^2  ~,
}
for $\Phi_\tx{ph} \propto z^X$,
we have
\eqn{\delta R = \frac{3d(d-1)}{(d+1) L_{d+1}^2} \Phi_\tx{ph}  = -\frac{3(d-1) \bar{R}}{(d+1)^2}  \Phi_\tx{ph} ~, \label{delta-R}}
\eqn{
\delta G_{zz} = d \times \delta G_{\sigma \sigma} = - \frac{d(d-1)(d-2)}{2(d+1)} \frac{\Phi_\tx{ph}}{z^2} = \frac{(d-1)(d-2) \bar{R}}{2 (d+1)} \frac{L_{d+1}^2}{z^2} \frac{\Phi_\tx{ph}}{d+1} ~,
}
where $\sigma=0$ or $i$ ($i=1,2,\cdots, d-1$).

\subsection{Tensor $h_{MN}=\delta g_{MN} -L^2_{d+1}\bar{\nabla}_M \bar{\nabla}_N \Theta_2$ \label{App-Linearized quantities-tensor}}
For the tensor perturbation $h_{MN}$ in \eqref{h}, which satisfies $\bar g^{MN} h_{MN}=0$ and $\bar{\nabla}^M h_{MN}=0$, the Ricci tensor is calculated as
\eqn{
\delta R_{MN} &= \bar{R}^K_{(M} h_{N)K} -\bar{R}^{L~K}_{~M~N} h_{KL} - \frac{1}{2} \bar{\Box} h_{MN} 
= -\frac{1}{2} \qty( \bar{\Box} + \frac{2(d+1)}{L_{d+1}^2} ) h_{MN} ~,
}
where 
the relations in (\ref{curvatures}) are used in the second quality.
We then  find
\eqn{
\delta R &= \bar{g}^{MN} \delta R_{MN} - h_{MN} \bar{R}^{MN} = 0 ~,\\
\delta G_{MN} &= \delta R_{MN} - \frac{\bar{R}}{2} h_{MN} =  -\frac{1}{2} \qty( \bar{\Box} - \frac{(d-2)(d+1)}{L_{d+1}^2} ) h_{MN} ~.
}
Since $\bar{\Box} h_{zz} =0$ and 
\eqn{
\bar{\Box} h_{\sigma \sigma} 
= -\frac{2}{L_{d+1}^2}  h_{\sigma \sigma} , \quad \sigma=0, i 
}
for $h_{MN} \propto z^{d-2}$,
we obtain $\delta G_{zz} = 0$ and
\eqn{
\delta G_{\sigma \sigma} =  \frac{d(d-1)}{2 L_{d+1}^2}  h_{\sigma \sigma} = - \frac{d-1}{d+1} \frac{\bar{R}}{2} h_{\sigma \sigma} ~. 
}

\section{$f(R)$ in Einstein frame \label{App-f(R) in Einstein frame}}
It is well known that $f(R)$ theory described by the Lagrangian (\ref{f(R)-Lagrangian}) is, at least classically, equivalent to the theory given by the following action with an auxiliary field $\chi$,
\eqn{
S= \frac{1}{2 \kappa} \int \dd^{d+1} x \sqrt{g} \qty( f(\chi) +  (R - \chi ) F(\chi) ) ~, \label{f(R) with auxiliary field}
}
where $F(\chi) := \p_\chi f(\chi)$,
since the EOM for $\chi$ reads $\chi = R$ as far as $\p_\chi F \ne 0$.

Reparametrizing $\chi$ with a scalar field $s$ as $F(\chi) = e^{ - (d-1) s }$,
one can move to the Einstein frame by Weyl transformation such that $\tilde{g}_{MN} = e^{ -2 s } g_{MN}$.
In the following, we put tildes on quantities associated with the new metric $\tilde{g}_{MN}$.
Since $ e^{2s} R = \tilde{R} -2d\tilde{\Box}s - d(d-1) \tilde{g}^{MN}\p_M s \p_N s$, up to surface term, the action (\ref{f(R) with auxiliary field}) is rewritten as
\eqn{
S= \frac{1}{2 \kappa} \int \dd^{d+1} x \sqrt{\tilde{g}} \qty( \tilde{R} -d (d-1) \tilde{g}^{MN} \p_M s \p_N s - 2 U(s) ) ~, \label{Einstein frame action}
}
where
\eqn{
 U(s) :=  e^{(d+1) s} \frac{ \chi F  - f }{2} = \frac{e^{2 s} \chi - e^{(d+1) s} f}{2} ~. 
}
The field $s$ is a scalar field with the potential function $U(s)$.

Let us define a constant field value $\bar{s}$
as the minimum of $U(s)$, which satisfies  
\eqn{
\p_s U = e^{(d+1)s} \qty(  \chi F - \frac{d+1}{2} f ) = 0 ~, \label{partial_s U}
}
which is equivalent to the condition (\ref{To determine Lambda}).
At $s=\bar{s}$, we find
\eqn{
\bar{\chi} \times e^{2 \bar{s}} = \frac{2(d+1)}{d-1}  U(\bar{s}) ~.
}
Given that $\chi = R$, we can identify the potential energy there as the cosmological constant in the Einstein frame
\eqn{ 
\tilde{\Lambda} := U(\bar{s})  =   e^{2 \bar{s}} \frac{\bar{R}(d-1)}{2(d+1)}
 = - \frac{d(d-1)}{2 \tilde{L}_{d+1}^2 } ~, \label{Einstein frame cosmological constant}
 }
 where $\tilde{L}_{d+1} := e^{-\bar{s}} L_{d+1}$.

We consider a small fluctuation $\delta s$ of the scalar field $s$ around $\bar{s}$: $s=\bar s+\delta s$. Since
\eqn{
 \p_s^2 U   = (d+1) \p_s U - (d-1) e^{2s} \qty(\frac{\p_s \chi }{2} +  \chi)  ~,
}
the mass of the canonically normalized scalar fluctuation $\tilde{s} := \delta s \sqrt{d(d-1)/\kappa}$ is obtained as
\eqn{
m_s^2 := \eval{\frac{\p_s^2 U}{d (d-1)}}_{s=\bar{s}} 
= - \frac{e^{2\bar{s}}}{d} \qty[ -\frac{d-1}{2} \frac{F}{\p_\chi F} +  \chi ]_{\chi =\bar{R}} ~, \label{m_s}
}
where we have used the relation $\p_s \chi = - (d-1) F / \p_\chi F $.
Since
\eqn{
e^{\delta s} = \qty( \frac{F(R)}{F(\bar{R})} )^{-\frac{1}{d-1}} = 1 - \frac{ 1 }{d-1} \frac{ \overline{\p^2 f} }{ \overline{\p f}} \delta R + \order{\delta R^2} ~, \label{delta-s}
}
we have $\delta s \propto \delta R \propto z^{d-2}$ from \eqref{delta-R} at the leading order.
This behavior becomes consistent with the linearlized EOM $(\bar{\tilde{\Box}} -m_s^2) \tilde{s} = 0$ if $m_s^2  = - 2 (d-2)/ \tilde{L}_{d+1}^2$
holds. Through (\ref{m_s}), this condition turns out to be  equivalent to the condition (\ref{second derivative}).

We now compute the metric perturbation in the Einstein frame.
By the Weyl transformation $\tilde{g}_{MN}=e^{-2s} g_{MN}$, it is essentially only the scalar mode that is affected.
The perturbation of the metric $\delta \tilde{g}_{MN}:=\tilde{g}_{MN}-\bar{\tilde{g}}_{MN}$ from the background AdS metric $\bar{\tilde{g}}_{MN}:= \delta_{MN}\tilde{L}_{d+1}/z^2$
is decomposed as
\eqn{
\delta \tilde{g}_{MN} = \frac{\bar{\tilde{g}}_{MN}}{d+1} \tilde{\Phi} + \tilde{h}_{MN} +  \tilde{L}_{d+1}^2 \qty(\bar{\tilde{\nabla}}_M \bar{\tilde{\nabla}}_N - \frac{\bar{\tilde{g}}_{MN}}{d+1} \bar{\tilde{\Box}} ) \tilde{\Theta} \label{tilde-delta-g}
}
where $\tilde{\Phi} := \Phi  - 2 (d+1) \delta s$
with $\Phi$ in (\ref{Phi}), $\tilde{h}_{MN} := e^{-2 \bar{s}}\times h_{MN}$
with $h_{MN}$ in (\ref{h}), and
$\tilde{\Theta} =  \Theta_1 + \Theta_2$
with $\Theta_1$ and $\Theta_2$ in (\ref{scalar}) and  (\ref{Theta_2}).
The physical scalar perturbation is defined by
$\tilde{\Phi}_\tx{ph} := \tilde{\Phi } -   \tilde{L}_{d+1}^2 \bar{\tilde{\Box}} \tilde{\Theta} = \Phi_\tx{ph} - 2 (d+1) \delta s $
with $\Phi_\tx{ph}$  in (\ref{scalar}).
Reading off $\delta s$ from (\ref{delta-s}) and plugging it the above, we obtain
\eqn{
\tilde{\Phi}_\tx{ph} = \Phi_\tx{ph} + \frac{2(d+1)}{d-1} \frac{ \overline{\p^2 f} }{ \overline{\p f}} \delta R = \qty(1 -\frac{6 \bar{R}}{d+1} \frac{ \overline{\p^2 f} }{ \overline{\p f}}  ) \Phi_\tx{ph} ~. \label{tilde-Phi_ph}
}
On the second equality, we have used (\ref{delta-R}).
With the condition (\ref{second derivative}), we find $\tilde{\Phi}_\tx{ph} = 0$.

The EOM following from (\ref{Einstein frame action}) for the gravitational field become
$\tilde{G}_{MN} + \tilde{\Lambda} \tilde{g}_{MN} - \tilde{T}_{MN}=0 $
\eqn{
\tilde{G}_{MN} + \tilde{\Lambda} \tilde{g}_{MN} - \tilde{T}_{MN}=0 ~
}
where the stress tensor is given by
\eqn{
\tilde{T}_{MN} := \p_M \tilde{s} \p_N \tilde{s} - \frac{\bar{g}_{MN} }{2} \qty( \bar{g}^{KL} \p_K \tilde{s} \p_L \tilde{s} + m_s^2 \tilde{s}^2 ) + \order{\tilde{s}^3}. \label{EMT in Einstein frame}
}
As in Sec.~\ref{Sec-Bulk theory},
we decompose it in the form of (\ref{EOM-decomposed}) and look only at $a=1$ with $b=-1,0$.
Since $\tilde{T}_{MN} \sim z^{2(d-3)}$ with $\tilde{s} \sim z^{d-2}$, the matter scalar field fluctuation does not contribute to the EOM at these orders.
Therefore,  the metric alone should solve the vacuum  Einstein equation.
First of all, the unperturbed  solution is given  by the AdS metric $\bar{\tilde{g}}_{MN}=\delta_{MN} \tilde{L}_{d+1}/z^2$ where $\tilde{L}_{d+1}$ is related to the cosmological constant $\tilde{\Lambda}$ in (\ref{Einstein frame cosmological constant}).
The leading order metric perturbation at $a=1$ with $b=-1$
is given by (\ref{tilde-Phi_ph}), which  vanishes due to the condition (\ref{second derivative}). 
Therefore the EOM is trivially satisfied at this order.
The actual leading order metric perturbation in the Einstein frame is the tensor perturbation $\tilde{h}_{MN}$,
which is the same as the Jordan frame tensor perturbation up to the factor $e^{-2 \bar{s}}$.
As mentioned below (\ref{h}), it solves the vacuum Einstein equation without any new condition, so that the $b=0$ part of EOM is satisfied.

Finally, we show that the metric (\ref{tilde-delta-g}) at this order is equivalent to the AdS black brane solution, given in
\eqref{metric_bb}.
Deviations of the black brane metric from the pure AdS spacetime are given in \eqref{delta-g_bb}.
On the other hand, the leading order perturbation of $\tilde{g}_{MN}$ in (\ref{tilde-delta-g}) with $\tilde{\Phi}_{\tx{ph}} =0$ becomes $\delta\tilde{g}_{MN}=\tilde{h}_{MN} +\tilde{L}_{d+1}^2 \bar{\tilde{\nabla}}_M \bar{\tilde{\nabla}}_N \tilde{\Theta}$, where $\tilde{h}_{MN}$ is given in \eqref{tilde-h} and $\tilde\Theta=\Theta_1+\Theta_2$.
With the identification in \eqref{Hawking-T},
it is easy to see that \eqref{delta-g_bb} and
\eqref{tilde-h} are equivalent by the gauge transformation as shown in \eqref{h-bb-equivalence}.
Therefore, (\ref{tilde-delta-g}) is equivalent to  the leading order perturbation in the AdS black brane with the Hawking temperature \eqref{Hawking-T}.

\section{GR with scalar field \label{App-Einstein+scalar}}
Let us consider a massive scalar field $\vs$ minimally coupled to Einstein gravity:
\eqn{\cl{L} = \frac{ R-2 \Lambda}{2 \kappa} - \frac{1}{2}\qty( g^{KL} \p_K \vs \p_L \vs + m^2 \vs^2)
\label{Einstein+scalar}}
with a negative cosmological constant $\Lambda = - d(d-1) / 2  L_{d+1}^2$.

The Einstein equation is given by
\eqn{
G_{MN} + \Lambda g_{MN}  = \kappa T_{MN} ~,
}
where 
\eqn{
T_{MN} := \p_M \vs \p_N \vs - \frac{g_{MN} }{2} \qty( g^{KL} \p_K \vs \p_L \vs + m^2 \vs^2 ) ~, \label{EMT}
}
while
EOM for the scalar field become $(\Box  -m^2 ) \vs = 0$.
At the zeroth order of $\kappa$, $g_{MN}$ becomes the AdS metric, and 
assuming $\vs$ depends only on $z$, we get a general solution on the background AdS spacetime as
\eqn{\vs(z) = \tilde{\vrho} \times z^{p_+} + \vrho \times z^{p_-} ~, \label{sol}}
where  $\vrho$ and $\tilde{\vrho}$ are arbitrary constants,
and 
\eqn{ p_\pm= \frac{d}{2} \pm \sqrt{ (d/2)^2 + L_{d+1}^2 m^2} ~.
}

At the first order of $\kappa$, 
the Einstein equation becomes
\eqn{\qty( \delta G_{MN} + \Lambda \delta g_{MN} ) = \kappa T_{MN} ~.}
In order for the behavior $T_{MN} \sim z^{2(p_\pm -1)}$ to match $\delta G_{MN} + \Lambda \delta g_{MN}  \sim z^{d-4}$ at the leading order as shown in (\ref{LHS}),
it is required that $p_- =(d-2)/2 = \Delta$ and $\tilde{\vrho}=0$ in the solution (\ref{sol}), which means that
\eqn{L_{d+1}^2 m^2 = -\frac{(d+2)(d-2)}{4} = \Delta (\Delta - d) ~.}
Since (\ref{EMT}) is evaluated as
\eqn{T_{zz} = \frac{d}{2} \times T_{\sigma \sigma} =\frac{d }{2} {d-2\over 2}  \vrho^2  z^{d-4} ~}
and the l.h.s. of the Einstein equation is given as (\ref{LHS}),
the minimally coupled scalar field can account for the perturbation $\delta g_{MN}$ in the UV limit with
the coefficient $\vrho$ determined by
\eqn{
 \vrho^2 = \frac{  T^{d-2} }{\kappa} \frac{   4  d (d-1) }{(d-2)^2 (d+2)} \frac{4 \zeta(d-2)}{\tx{B}(\Delta, d-\Delta)}  ~.
}

According to the relation (\ref{GKPW}), the bulk degree of freedom $\propto z^{\Delta}$ has corresponding operator in the boundary theory with the conformal dimension $\Delta$,
whose unique candidate of such a operator is $\vp^a$.
However, its thermal expectation value vanishes 
due to the unbroken $O(N)$ symmetry, even though the conformal symmetry is broken by the temperature. 
Therefore, the relation (\ref{bulk-excitation}) is not satisfied for $\vp^a$,
and we thus conclude that the bulk theory described by (\ref{Einstein+scalar}) can not be dual to the boundary theory in (\ref{free_action}).


\bibliographystyle{apsrev4-2}
\bibliography{BulkReconstruction}

\end{document}